\documentclass[useAMS,usenatbib,epsfig]{mn2e} 
\ifx\pdfoutput\undefined

\usepackage[pdftex]{graphicx}
\else
\usepackage[pdftex]{graphicx}
\usepackage{epstopdf}
\fi

\usepackage{algorithmic}
\usepackage{natbib}
\usepackage{amsmath,amsfonts}
\usepackage{epsfig}
\usepackage{times}
\def\beq{\begin{equation}}
\def\eeq{\end{equation}}

\title[COSMOSOMAS 10--12 GHz Data.]{COSMOSOMAS Observations of the CMB and Galactic  Foregrounds at 11 GHz: 
Evidence for anomalous microwave emission  at high Galactic Latitude.}
\author[S. R. Hildebrandt et al.]
{S. R. Hildebrandt$^1$, 
R. Rebolo$^{1,2}$,
J. A. Rubi\~{n}o-Mart\'in,$^1$, 
R. A. Watson$^{1,3}$, 
\newauthor C. M. Guti\'errez$^1$, 
R. J. Hoyland$^1$ and 
E. S. Battistelli$^{1,\star}$ \\ \\
$^1$Instituto de Astrof\'\i sica de Canarias, 38200 La Laguna,
  Tenerife, Spain.\\
$^2$ Consejo Superior de Investigaciones Cient\'\i ficas, Spain. \\
$^3$ Jodrell Bank Observatory, University of Manchester,
  Macclesfield, Cheshire SK11 9DL, UK.\\
$^{\star}$ Present address: University of British Columbia
Department of Physics and Astronomy. 6224 Agricultural Road
Vancouver, B.C. Canada V6T 1Z1.}
\pagerange{\pageref{firstpage}--\pageref{lastpage}}
\pubyear{2004}
\begin{document}

\date{Accepted ---; received ---; in original form \today}
\pagerange{\pageref{firstpage}--\pageref{lastpage}}
\pubyear{2005}
\maketitle

\label{firstpage}

\begin{abstract}

 We present  observations with the  new 11 GHz radiometer  of the COSMOSOMAS experiment at 
the Teide Observatory (Tenerife). The sky region between
 $ 0^{\circ} \le RA \le 360^{\circ} $ and $ 26^{\circ} \le DEC \le 49^{\circ} $ 
(ca. 6500 square degrees) was observed with   an angular resolution  of $ 0.9^\circ$. 
Two orthogonal independent channels in the receiving system measured total 
power signals from  linear polarizations
with a 2 GHz bandwidth. Maps with an average sensitivity  of
50 $\mu$K per beam have been obtained for each channel.  
At high Galactic latitude ($|b|>30^{\circ}$) the 11 GHz data
are found to contain  the expected cosmic microwave background as well as 
extragalactic radiosources, galactic synchrotron and free-free emission, and  a dust-correlated component 
which is very likely of galactic origin. At the angular scales allowed by the window
 function of the experiment, the dust-correlated component  presents an 
 amplitude $\Delta T \sim$ 9--13 $\mu$K while the  CMB signal is of order 27 $\mu$K. The spectral
 behaviour of the dust-correlated 
signal is examined in the light of  previous COSMOSOMAS data at 
13-17 GHz and WMAP data at 22-94 GHz in the same sky region. 
We detect a flattening  in the spectral index of this  signal below 20 GHz which rules out synchrotron
radiation as being responsible for the emission. This anomalous dust emission 
can be  described by a combination of free-free emission and 
spinning dust models with a flux density peaking  around 20 GHz.

\end{abstract}
\section{Introduction}

Since 1984, the Teide Observatory  has hosted experiments  to observe 
the  cosmic microwave background at 10-33 GHz on angular scales from 
degrees to  arcminutes. Located at this observatory,  the COSMOSOMAS 
experiment consists of  two radiotelescopes 
designed to acquire sensitive observations of the microwave sky at 
frequencies in the range 10-17 GHz with an angular 
resolution of $\sim$ 1 degree. It was conceived 
as a  high resolution extension of the  Tenerife experiment 
\citep{davies96,hancock/etal:1997,gut2000} with the goal to provide 
 10-17 GHz ground-based data of angular  resolution comparable to  
the lowest frequency channel of the Wilkinson Microwave Anisotropy Probe, WMAP 
\citep{bennet/etal:2003a}. Observations in this frequency range can contribute 
significantly to a better understanding  of the foreground emission and to improve
the  limits imposed on cosmological parameters  by present and future CMB experiments. 
This appears  particularly relevant for missions like  
Planck\footnote{{\tt http://www.rssd.esa.int/index.php?project=PLANCK}}
which are planned to achieve a sensitivity  of order 1 $\mu$K per resolution element 
and infer cosmological  parameters with accuracy better than 1\%.

There is also an increasing interest in sky maps  at few tens of  GHz to 
elucidate the presence of a new microwave emission process different to the three 
classical Galactic foregrounds, namely, synchrotron, free-free and thermal  
dust emission. First detected as a dust-correlated signal  
in the COBE/DMR data (Kogut et al. 1996), this microwave emission was initially interpreted as 
due to free-free radiation. However, statistical detections of dust 
correlated signals by experiments like CBI \citep{mason01},  
Saaskaton  \citep{deOliveira-Costa/etal:1997},  OVRO  
\citep{leitch/etal:1997},  the Tenerife experiment 
\citep{angelica99, angelica04}  and also in the Green Bank Galactic Plane 
Survey  (4.85 GHz) \citep{finkbeiner/etal:2004}  strongly suggested that 
free-free emission cannot be the only underlying process.  Draine and Lazarian (1998a,b) 
alternatively proposed electric dipole radiation from spinning 
particles and  magnetic dipole emission as
mechanisms behind this anomalous microwave emission.

A dust correlated signal has also been detected in the WMAP data 
(Lagache 2003, Bennet et al. 2003a) but its interpretation is still a 
matter of debate.  Bennet et al. 2003a argue that all foreground emission 
observed at WMAP frequencies can be explained in terms of the three classical 
foreground components. However, de Oliveira-Costa et al. (2004) 
show that
the synchrotron template at 22 GHz derived by the WMAP team 
under the previous assumption correlates with the 10 and 15 GHz  
Tenerife data in a manner incompatible with the   
frequency behaviour of synchrotron emission. 
More recently, Fern\'{a}ndez-Cerezo et al. (2006) using 
high Galactic latitude observations at 13-17 GHz obtained with the first radiometer (COSMO15) 
of the COSMOSOMAS experiment (over an area of more than 3000  ${\rm deg}^2$)  
find evidence that below 20 GHz the spectrum of the dust 
correlated signal turns over just as predicted by 
spinning dust emission. Unambiguous evidence for 
anomalous microwave emission in individual astronomical objects has been 
found by the COSMOSOMAS experiment (Watson et al. 2005) in the Perseus molecular complex and by CBI 
in the dark cloud LDN1622 (Casassus et al. 2006).

In this paper,  we present new  results of the COSMOSOMAS experiment\footnote{{\tt http://www.iac.es/project/cmb/cosmosomas} of  the Instituto de Astrof\'{\i}sica de Canarias.} obtained 
 with the 11 GHz radiometer, hereafter referred as COSMO11. 
These observations cover a sky region of  6500 ${\rm deg}^2$ with 
an angular resolution of $\sim1^{\circ}$ and sufficient sensitivity 
to explore further the behaviour of the anomalous microwave emission 
and other foregrounds at high Galactic latitudes.  While we  
find cosmic microwave background and extragalagtic radiosources  
dominate the temperature fluctuations at 11 GHz in our beam scale, we also detect 
a highly significant dust-correlated signal at this frequency which is
compared with previous results in the same region using COSMO15 
 and WMAP data.

\section{The COSMO11 radiometer}

 The COSMOSOMAS experiment consists of two
independent  similar radiotelescopes (COSMO15 and COSMO11) aimed to obtain 
measurements of the  microwave radiation in the frequency range 10-17 GHz with average resolution
of 0.9$^\circ$ (FWHM). The experiment is sited at
Teide Observatory (2400m altitude, Tenerife) and details of the observational strategy can be found in
\citep{gallegos/etal:2001}. Essentially, each of the two instruments
performs daily observations of a strip
in the sky with diameter 20$^\circ$ in DEC, centred at the meridian.
Earth rotation produces a map with complete RA coverage. An area equivalent to $\sim 14\% $ 
of the whole sky is then covered daily with a typical
sensitivity of 1.0--2.0 mK per pixel, or roughly 0.3--0.7 mK per beam, after
removal of most atmospheric contamination (see Sect.~\ref{sec-day}).

COSMO11 started scientific operation at the end of 2003, several years later
than the higher frequency  COSMO15, however its good overall performance
has resulted in similar sensitivities to COSMO15 in a much
shorter time scale. In this work, we present COSMO11 and
analyze data taken from  November 2003 to 10$^{\rm th}$ June 2005.
By stacking the  best first 150 good observing days, COSMO11 achieves a
comparable sensitivity,  40--50 $\mu$K per beam, to that
of COSMO15 data \citep{cerezo/etal:2006}.

The COSMO11 radiometer was equipped with two 
similar receivers, working at almost identical frequencies, but 
sensitive to orthogonalpolarizations in order to provide
a better control of systematics and explore potential 
polarization measurements.
 
\subsection{Experimental layout}
\label{sec-exp}

\begin{figure*}
\begin{center}
\includegraphics[width=15cm,height=12cm,angle=0]{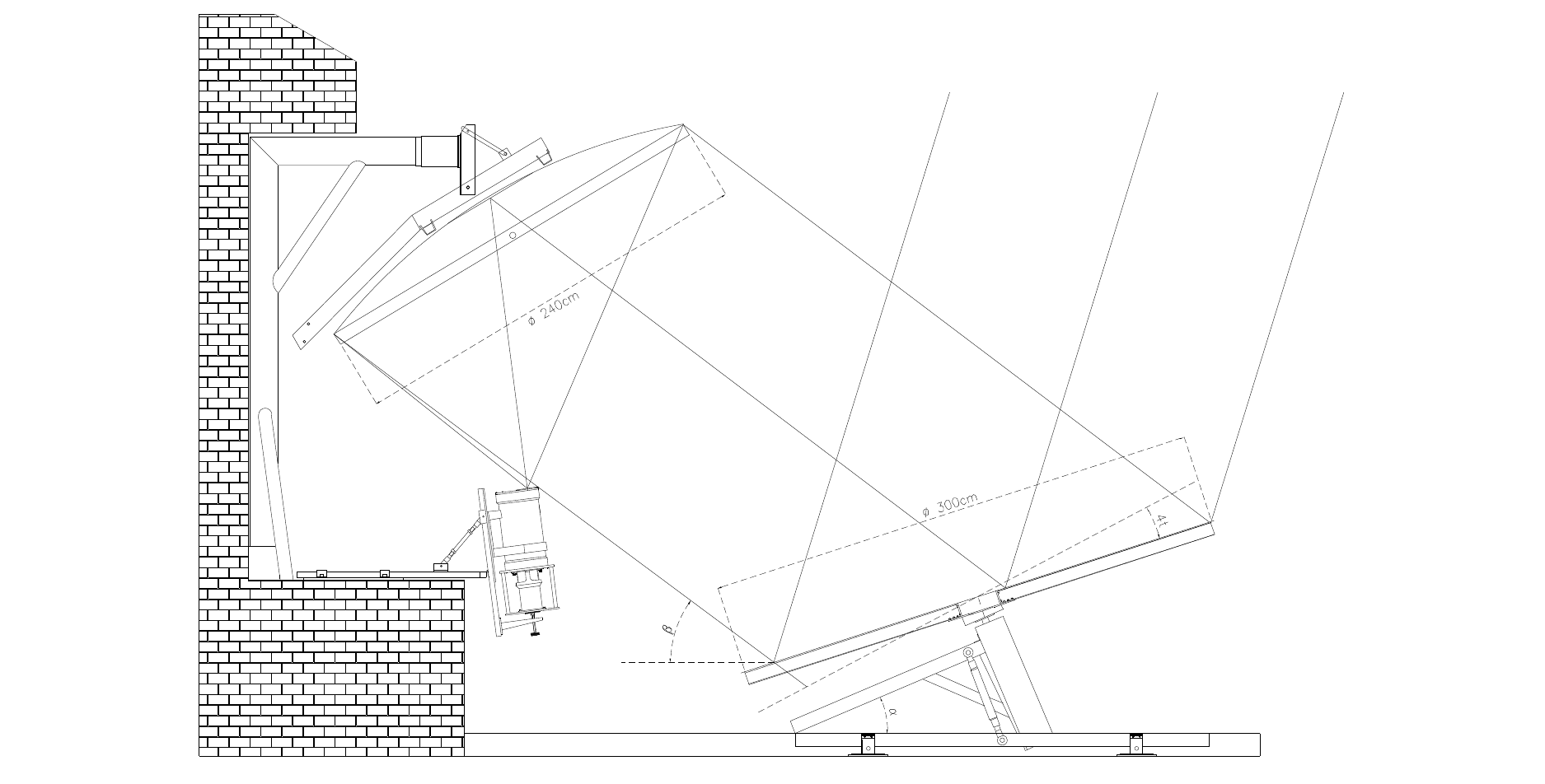}
\caption{Optical system and mechanical structure of COSMO11.}
\label{fig-cosmo11}
\end{center}
\end{figure*}

The optical and mechanical structure of the COSMO11 experiment (see Fig.~\ref{fig-cosmo11}) is very similar to that of COSMO15. The main  differences are the size of the circular, primary, flat,
mirror of
3.0 m (instead of 2.5 m) and  the
aperture of the parabolic, secondary, mirror which is
2.4 m instead of  1.8 m. Both instruments were  designed to achieve an angular resolution
of 0.9$^\circ$. A metal block with a 5$^\circ$ opening angle tilts the main, flat mirror. In this way as the mirror spins (at $\sim $1 Hz), the observed path in the sky describes a circle of 20$^\circ$ of diameter.
During the period of observations relevant to this paper the inclination of the  primary mirror has been changed twice while  the position of the parabolic dish and the tilt of the primary mirror were kept fixed. A change
of the inclination of the primary mirror causes a displacement  in DEC
of the center of the ring described by the beam on  the sky. In
Tab.~\ref{table-configurations11} we give a summary of the values for the main  geometrical
parameters.  The setup adopted for the observations led
to data acquisition from declination 23.80$^{\circ}$ to  50.80$^{\circ}$.

\begin{table}
\caption{Summary of geometric configurations and
observed declinations (complete in RA) in this work.
Tilt has always been  5$^{\circ}$. Units in degrees.}
\vspace{0.3cm}
\begin{tabular}{l c c c c}
Period & $\alpha $ & $\beta$ &
         $ {\rm DEC}_{\rm min}$  & $ {\rm DEC}_{\rm max} $\\
\hline
\hline
01/11/03-12/05/04  &  26.00 & 64.50  &  30.80 & 50.80 \\
23/05/04-10/06/05  &  29.50 & 64.50  &  23.80 & 43.80 \\
\hline
\end{tabular}
\label{table-configurations11}
\end{table}

\subsection{Receivers}
\label{sec-receivers}
\begin{figure}
\begin{center}
\includegraphics[width=8cm,height=11cm,angle=0]{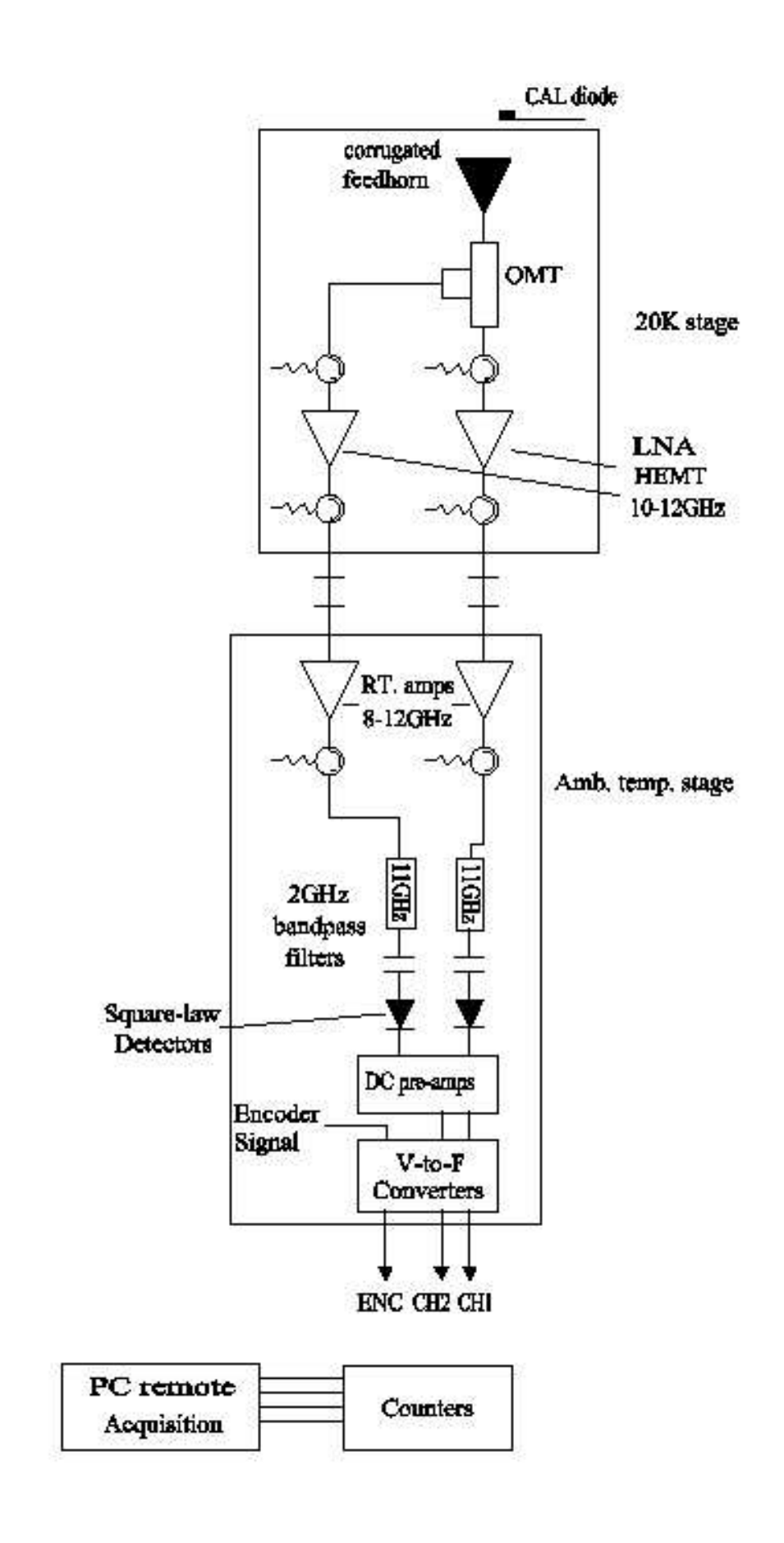}
\caption{Receiver system of COSMO11. Two polarizations of total power signal at $\sim$ 10.9 GHz are
observed.}
\label{fig-receivers11}
\end{center}
\end{figure}

The COSMO11 receiver is a polarimeter, designed to receive orthogonal linear polarisations through 2 total power channels  (see Fig.~\ref{fig-receivers11}).

\begin{figure}
\begin{center}
\includegraphics[width=7.5cm,angle=0]{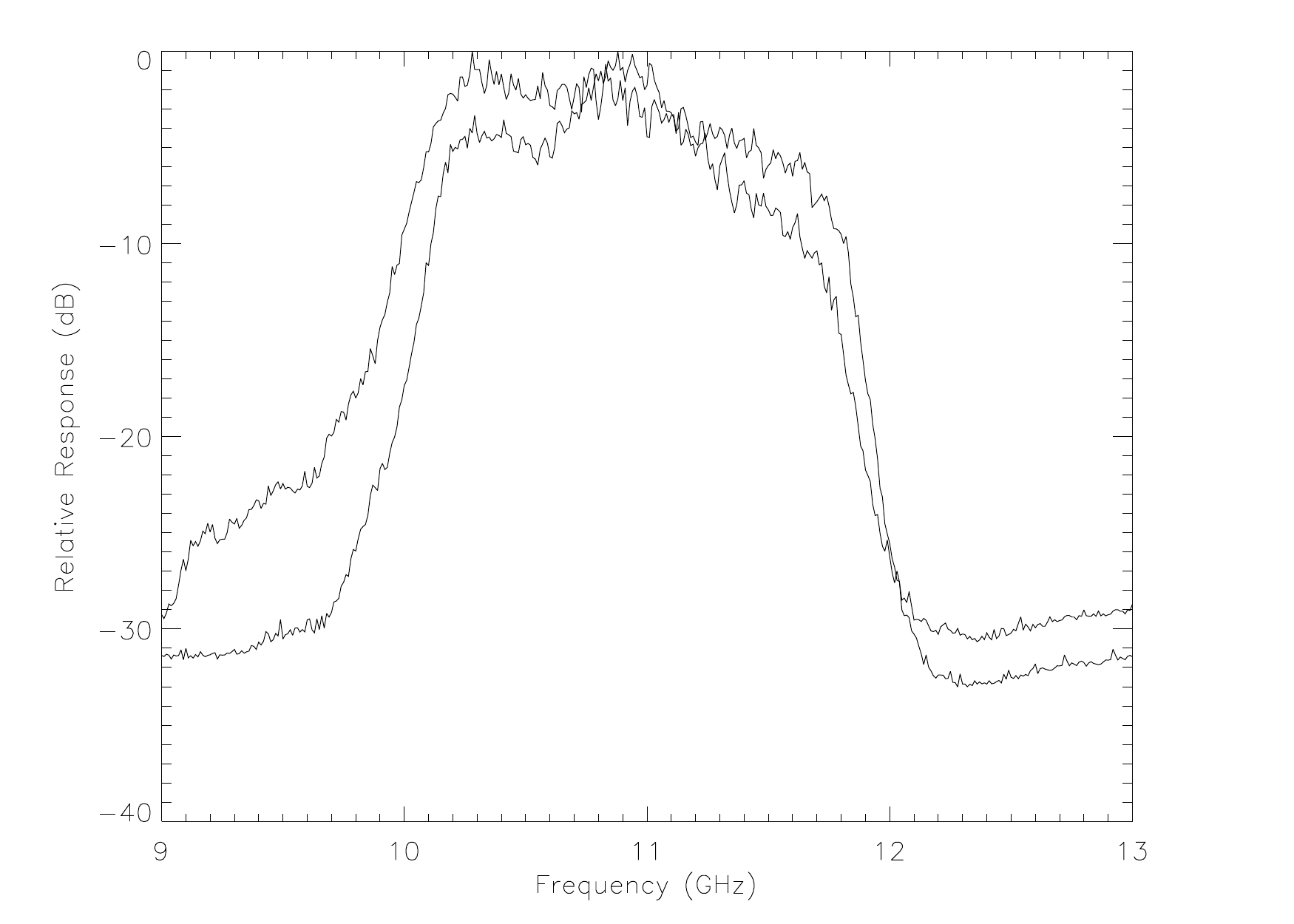}
\caption{Relative spectral responses of the two filter band passes in COSMO11.}
\label{fig-filters11}
\end{center}
\end{figure}

The whole front-end of the receiver is housed in a cryostat and cooled to 15-20K by a dual-staged helium gas refrigerator. The phase centre of the feedhorn coincides with the focal plane of the offset parabola. This provides the alignment of the cryostat. The cryostat window is a sandwich of thin Mylar film and high density polypropylene foam. Secondary calibration is provided by a temperature controlled broadband noise source which enters the beam through a side lobe. This coupling is obtained by placing the calibration probe close to the cryostat window but not in the main beam.
The feed horn is a conical corrugated design for maximum efficiency and minimum side lobes. 
The feed horn is followed by an orthomode junction. Both are  designed by ERA technology Ltd and  
are matched to the Channel master 2.4m offset parabolic antenna. The two
 linear polar channels feed two similar 10-12GHz cryogenically cooled amplifiers from 
Berkshire Technologies Inc. The noise temperatures are close to 10K and the gain is 25dB minimum. 
The complete total power channels form direct detection chains whereby a matched square-law detector is used to detect the microwave signal without first mixing the signal down to an intermediate frequency. The receiver response just before entering the detectors is shown in Fig.~\ref{fig-filters11}. The mean, weighted, frequency defined as:
\begin{equation}
{\bar \nu} := {{\large \int}_{\nu_{min}}^{\nu_{max}} \nu \,\rho(\nu) \, d\nu},
\end{equation}
where $ \rho (\nu) $ is the normalized (by setting it's integral to unity)
linear spectral response of the filters,
yields 10.89 GHz for channel 1 and 10.87 GHz for channel 2.
The overall 1/f knee frequency is about 3Hz, which is faster than the spinning flat plate frequency of 1Hz. The 1/f is removed from the data stream by eliminating sufficient Fourier harmonics. See section 2.3.
The rest of the receiving system is  analogous to that of COSMO15
\citep{cerezo/etal:2006}.

\subsection{Receiver orientation and scanning strategy}

In this work, we only deal with data obtained for the primary  configuration with channel 1 and 2 polarizations oriented
North-South and East-West, respectively, see Tab.~\ref{table-polarization}. 
In order to perform polarization studies, it is necessary to rotate the
receiver system and take data with a second orientation, e.g., 
with a 45$^{\circ}$ relative rotation
with respect to  the primary configuration. Observations in this second 
configuration are being currently conducted
and polarization measurements of the Galactic foregrounds and the cosmic 
microwave background will be considered in a future paper, although a 
faint polarized emission has been already measured  
in the
Perseus molecular complex  with the first data 
obtained so far \citep{battistelli/etal:2006}.

 In order to minimize the effect of systematics in the data set, we also 
 rotated the cryostat by 90$^\circ$, which  essentially produces an 
 exchange in the receivers of channel 1 and 2.  

\begin{table}
\caption{Summary of the receiver configurations. The
orientations referred to local meridian. 
Useful days are also given for Channel 1 and 2, respectively.}
\vspace{0.3cm}
\begin{tabular}{l c c c}
Period &  Channel 1 & Channel 2 & Useful days\\
\hline
\hline
01/11/03-10/06/05  & North-South  & East-West  & 135, 140 \\
\hline
\end{tabular}
\label{table-polarization}
\end{table}
 
\begin{figure*}
\begin{center}
\includegraphics[width=15cm,angle=0]{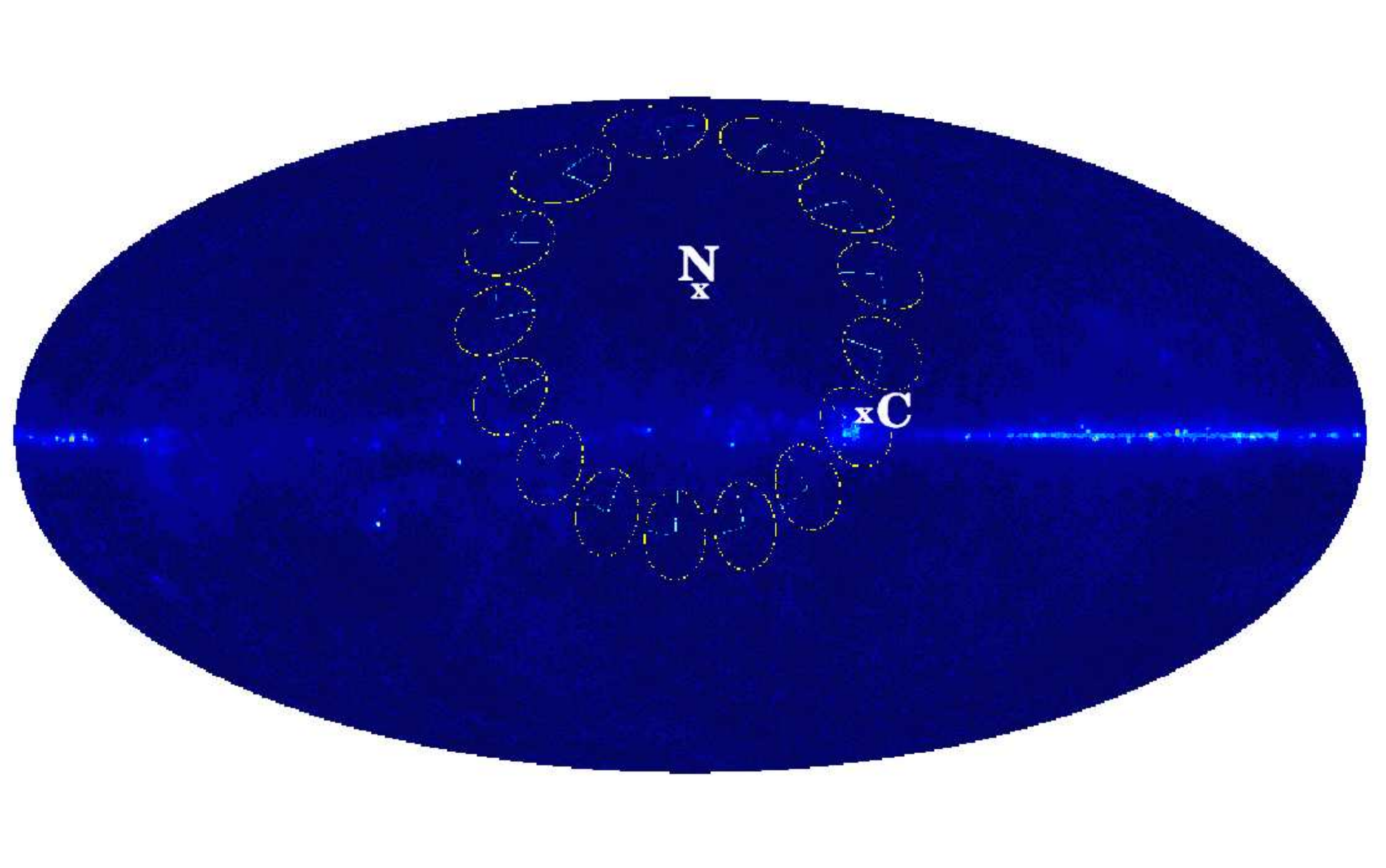}
\caption{COSMO11 scanning strategy together with the two polarization
directions observed in Galactic coordinates.  The image used for the
sky is that provided by  WMAP for its lowest frequency channel
(K-band, 22.5 GHz), which is the closest to COSMO11 observed
frequencies. Top: image centered at Galactic longitude
0$^{\circ}$. The Celestial north pole is again marked with an ``N''.
Cygnus A complex is labelled with a ``C''. HEALPix pixelisation scheme is used,
\citep{healpix}.}
\label{fig-pol_ortho}
\end{center}
\end{figure*}

In Fig.~\ref{fig-pol_ortho}, the scanning
strategy of COSMO11 is shown. The circles in RA and DEC, the 
sky coverage due to Earth rotation, and the two polarizations,
pointing towards the Celestial north pole: one North-South; the other,
East-West. Since we are also interested in Galactic foreground emission,
we also give a figure of the scanning strategy in Galactic coordinates.
The sky image is that of the lowest frequency WMAP channel, i.e. K band.
One can readily see  the crossing of COSMOSOMAS rings with
the Galactic plane, both center and anti center regions, and the high
Galactic latitude region  observed.  We also note the position of  Cygnus A, our main calibrator. 

\subsection{Data acquisition}
\label{sec-data}
The COSMO11 data acquisition procedures are similar to those used in  COSMO15. Data are
sampled over 4000 $\mu$s, leaving a blanking time of order 400 $\mu$s.
A Fast Fourier Transformation is applied to the data for each mirror
cycle, keeping 106 harmonics.  In practice, we find  that for the construction of the daily 
maps, it is enough to consider up to the $45^{\rm th}$
harmonic; higher harmonics do not contain astronomical information.
Then, the data gathered every $\sim $ 30s are stacked
together, yielding a mean value for each of the 106 harmonics and
their corresponding standard deviations. We note that in 30 s the
apparent motion of the sky corresponds to $\sim$ 0.12$^\circ$. Since
the beam size is around 0.9$^\circ$, there is no problem in averaging 
data every $\sim$ 30 s. This is done mainly to save memory and allow
faster data processing. Nevertheless,  we store independently the  data obtained for
each spin of the mirror every  $\sim$ 1 s  mainly for the technical monitoring of 
the data acquisition system.  Every second, a diode signal of 2 K is introduced in the feed horn.
This allows to measure changes in the receiver gain. Each day of data is saved into a FITS binary file.

Every 30 s a stacked data set, called here a {\em scan} is stored. Each scan is 
comprised of 275 samples. In this way, each sample corresponds to an angular size of order
$  \pi D /275 $, where $D$ is the diameter of the circle subtended by the
COSMO11 beam on the sky. From Sect.~\ref{sec-exp}, $D$ is 
$\sim 20^\circ$ and therefore the angular size of each sampling index 
is $\sim 0.23^\circ$. Summarizing, the angular size of the data bin is of order
$ 0.12^\circ \times 0.23^\circ $, which is appropriate for the beam
angular resolution of COSMO11 (eventually, for the astronomical map we will choose
a pixelization of ${1/3}^\circ \times {1/ 3}^\circ $).

\section{Data processing}
\label{sec-c11-processing}
In this section, we describe the data analysis of COSMO11 observations, from 
the  daily map making procedure and calibration to the  cumulative maps corresponding  
to  a period of approximately 110 equivalent  observing days for  each channel.

\subsection{Daily processing}
\label{sec-day}

 The first step is the correction of gain fluctuations in the data.  To that end, we 
divide the counts measured in  each scan by the value of the 2 K calibration diode
corresponding to  that scan. A first clipping of bad data is performed on the scans consisting
of rejection of those with  saturated signals,  no signal, spikes and/or glitches. The result is a  
sequence of scans with total power signal in acceptable ranges.

\begin{figure}
\begin{center}
\includegraphics[width=7.5cm,angle=0]{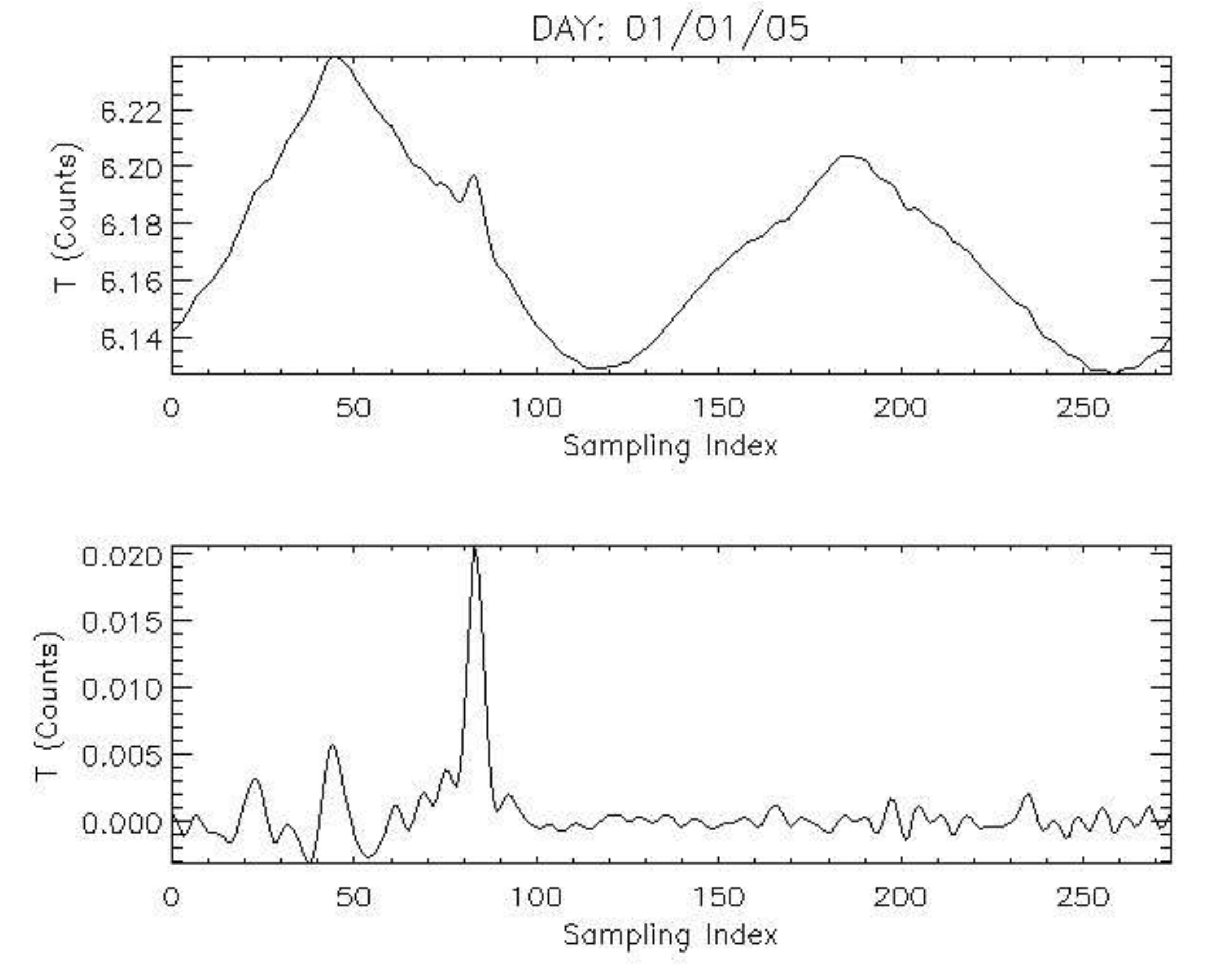}
\caption{Typical  scan obtained with  COSMO11. Up: atmospheric emission appears clearly as 
a long-scale modulation. The peak close to sampling index 85 corresponds to 
the observation of the  Cygnus A region. Down: total power after subtraction of the first seven Fourier harmonics. Structures to the left side of the Cygnus A 
region are associated to nearby
 galactic plane regions.}
\label{fig-cleaning11}
\end{center}
\end{figure}

In addition to instrumental noise, our data are affected by atmospheric 
emission which causes a modulation in the total power every spin of the primary mirror
(see Fig.~\ref{fig-cleaning11}). After probing 
some techniques to remove this  atmospheric modulation, we arrived to 
the conclusion that subtraction of a Fourier fit to  {\it each} scan is 
a suitable  and efficient procedure. In this way, 
we also  remove the main part of $1/f$-like  noise. At 11 GHz,
it suffices to remove some low order Fourier terms in order to get 
useful data (see later, Fig.~\ref{fig-galactic-center-1-3}). However,
for the sake of a direct comparison with the previous  COSMO15 data, notably more affected by atmospheric emission than the current experiment, 
we choose here the same number of Fourier terms to be removed in each scan, i.e., 
a Fourier series fit of 7$^{\rm th} $ order (i.e., a constant term plus seven $\rm sin $ and $\rm cos$ terms). 
The fit of any scan is thus independent from the rest. In the Fourier series fit 
to each scan, we mask any single data sample within a 
radius of 0.4 $^\circ$  of  a signal exceeding 4 times the standard deviation of
the scan data. This is done  to minimize the effect of bright sources on the fitted curve.  After this individual scan cleaning a second clipping is applied consisting of rejection of scans with standard deviation  20 times larger than the mean standard deviation of the three preceding and three following scans.

 After this processing has been executed, we still find a much weaker daily modulation of the data taken at the same sky position (same line of sight or same sample index ). This modulation is possibly associated to day-night atmospheric temperature variations.  We correct it removing the  first four Fourier terms in the fit to the whole day-night data set in any sky position. Here again, data in a 0.4 $^\circ$  interval centred in points which exceed the day-night mean value in 4$\sigma $ are masked in the Fourier series fit.
The correction applied to the day-night modulation modifies the window function of the experiment \citep{ga01}. Because of the masking of the bright sources during the fitting procedure applied to the day-night modulation, the effect on point sources is very small, however CMB, very weak sources and diffuse foregrounds are affected and will 
be taken into account when generating adequate templates for
cross-correlation with the experiment data. 

\subsubsection{Daily map making}
\label{sec-daymap}
\begin{figure}
\begin{center}
\includegraphics[width=9.5cm,angle=0]{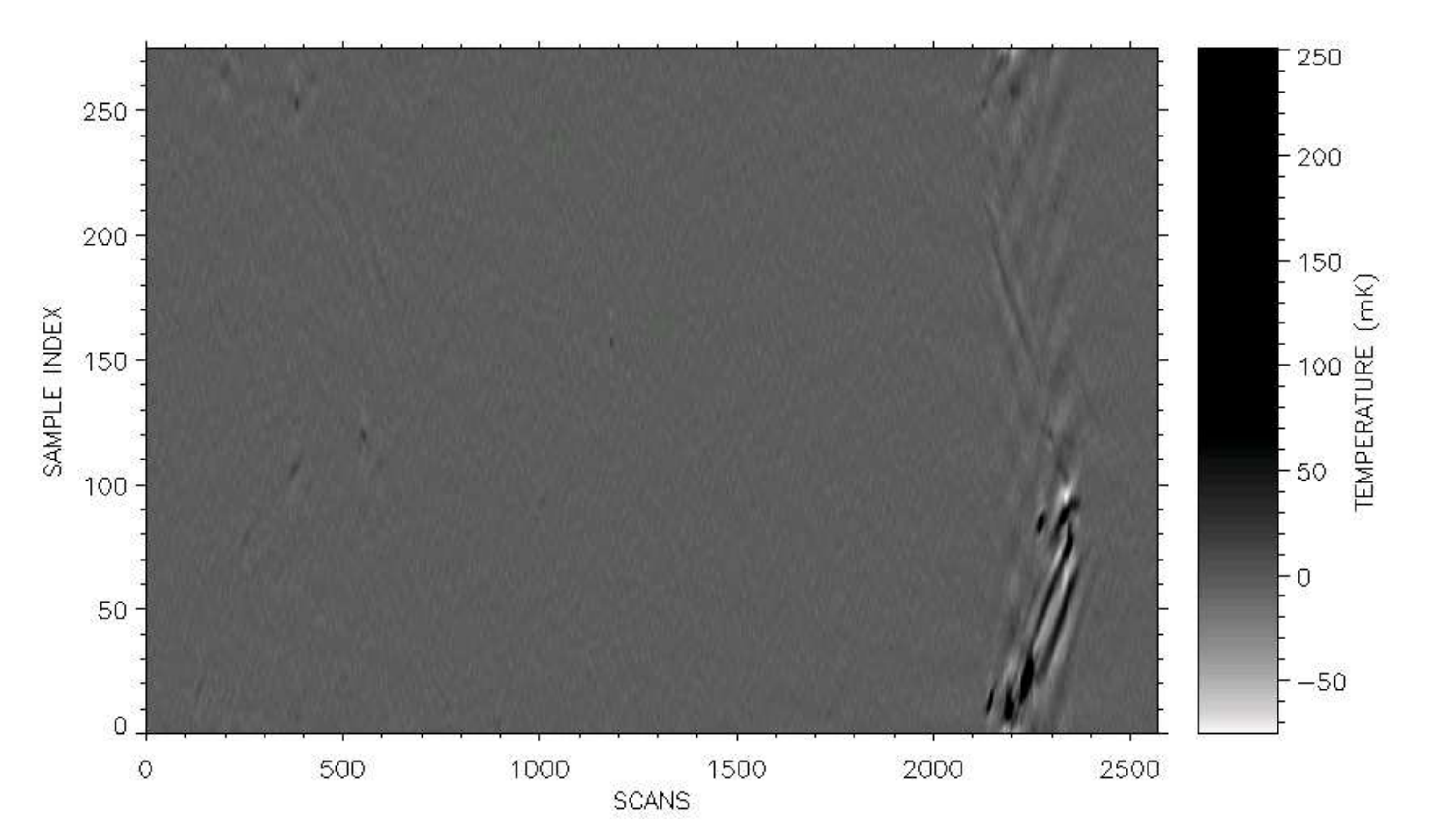}
\caption{Scan sequence for a typical day in the COSMO11 data. Scans are 
cleaned. Notice the duplication of the observed sky region. The Galactic Cygnus region and Cygnus A are clearly observed near scan number 
2000. The brightest sources seen along the first scans are the complex regions towards the  Galactic  anti-center.} 
\label{fig-seqcleaning11}
\end{center}
\end{figure}

Due to the scanning strategy of COSMO11, 
each astronomical source  is observed twice (except in the upper and bottom 
sections of the scan). Fig.~\ref{fig-seqcleaning11} shows the sequence
 of scans divided into its 275 sampling sections (vertical axis) in terms of the 
scan number (time). The two point-like sources placed at approximately 
(2150,12) 
and (2250,85) correspond to the first and second crossing of Cygnus A with the beam 
of the experiment. Other sources are also seen daily, but
with insufficient signal/noise ratio  to be used as primary calibrators. 
For example, GB6 0319+4130 (3C84) can be seen at around (200,270). This source  is known to show variability. From the 1980s
 until the epoch of the COSMOSOMAS observations, its flux has decreased from $\sim$ 
50 Jy to $\sim $ 20 Jy. Our  observations are in agreement with the flux  extrapolated 
from the WMAP K band measurement ($S = 11.1 \pm 0.1 \,$Jy) and  the spectral index 
 published for this source  by the WMAP team ($\alpha = -0.8 \pm 0.1)$.
 The other bright sources seen in the map are complex structures in the Galactic plane
towards the anti center.

This double measurement of any sky region is then used in 
the construction of a daily map.
First, knowing the local time of the entrance and exit of a known point-like
source, say Cygnus A, we compute their hour angle (HA). Next, we use the 
values of the declination of the source and also the direct measure of the 
inclination of the main, flat, mirror ($\alpha$) and of the tilt angle
($ \theta $) as starting values to compute optimal values for the geometrical
parameters that make to coincide  the derived and observed HA. 

\begin{figure}
\begin{center}
\includegraphics[width=9.5cm,angle=0]{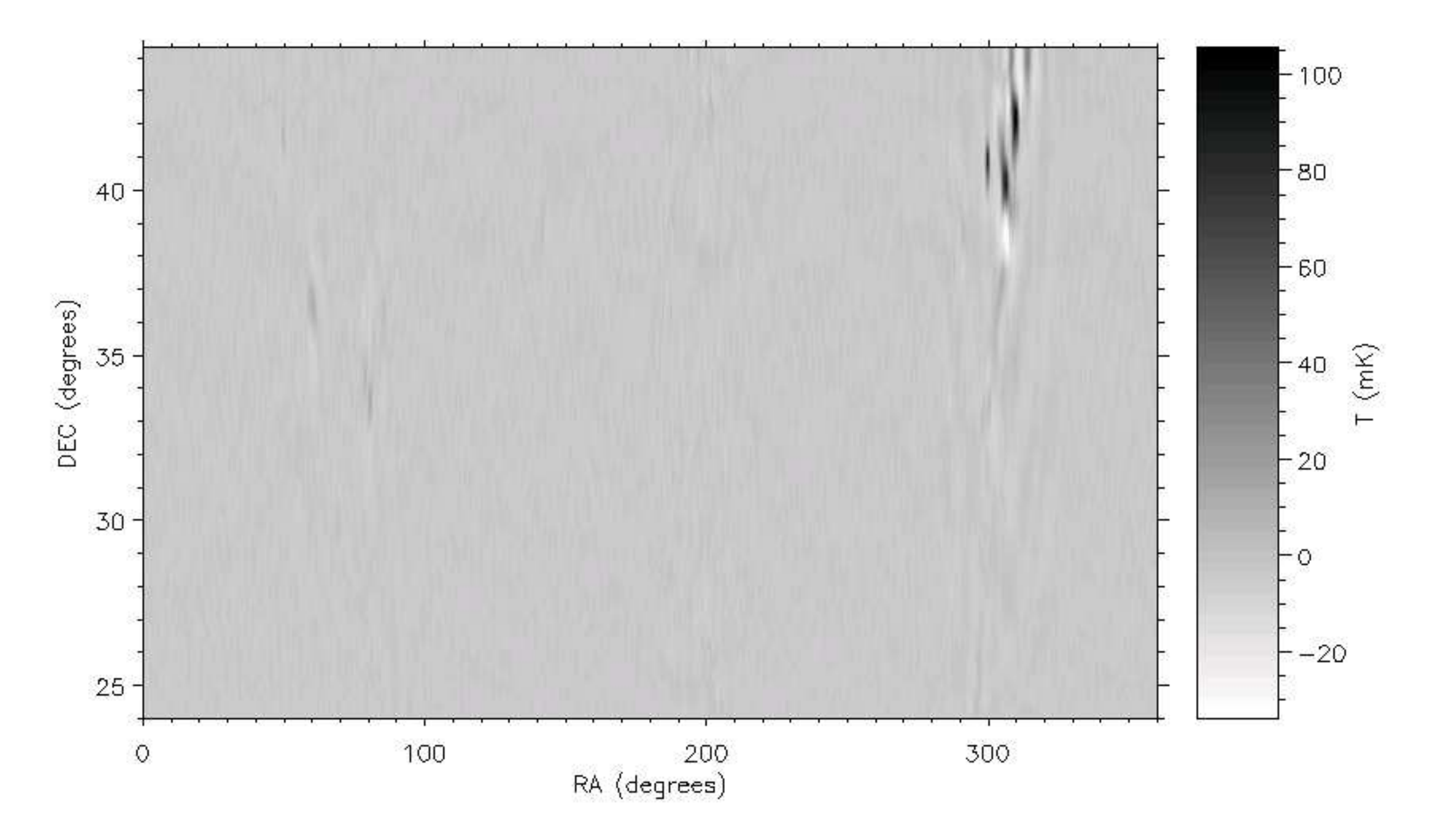}
\caption{Typical daily map. 
The Galactic Plane  crossing  is clearly seen, 
including the Cygnus A complex. Also clearly 
detected is the crossing towards the anti center region.}
\label{fig-daymap11}
\end{center}
\end{figure}

 Once, $ \alpha $, $ \beta $ and $ \theta $ are inferred, we construct
a daily map by projecting each of the observed pixels into a rectangular grid
with RA in its horizontal axis and DEC in its vertical one 
(see Fig.~\ref{fig-daymap11}). The chosen pixelization is $1/3^\circ
\times 1/3^\circ $, which is adequate for the  angular size of the sampling
index and for the size of the beam. Due to the range of observed
declination (20$^\circ$), this method is appropriate and we do not resort here to more refined
 methods as the Healpix pixelization. 

\subsubsection{Daily flux/temperature calibration}
\label{sec-calibration}

\begin{figure}
\begin{center}
\includegraphics[width=6cm,angle=0]{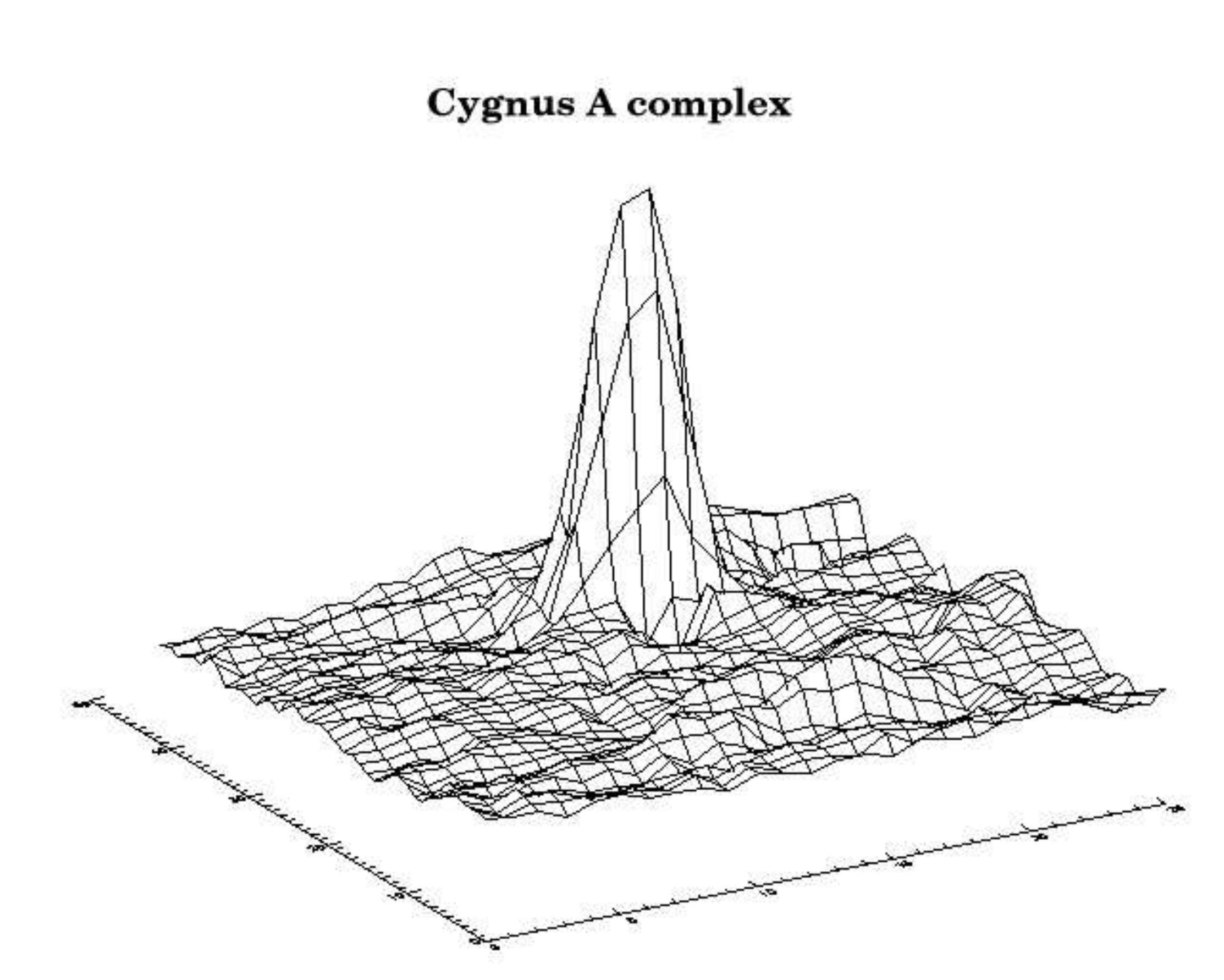}
\caption{Cygnus A as recorded in a \it{single} day of observation by COSMO11.}
\label{fig-cyga11}
\end{center}
\end{figure}

 Each daily set of data is calibrated in flux or antenna temperature with
observations of the brightest stable point-like radio source in our maps: Cygnus A
(see Fig.~\ref{fig-daymap11}). In Fig.~\ref{fig-cyga11} one of the  daily images of Cygnus A 
observed by  COSMO11 is plotted.

We use the empirical model by ~\citep{baars/etal:77} 
to derive the expected flux density ($S$) 
from Cygnus A. For our frequency range it gives
\beq
\label{eq-S}
\log S^{B}_{\,\rm{Cyg}} [\rm{Jy}] = a + b \, \log\nu[{\rm MHz}],
\eeq
where $ a = 7.161 \pm 0.053 $ and $ b = -1.244 \pm 0.014 $ which
leads to a flux of Cygnus A of $ S_{\rm Cyg}(10-12 {\rm GHz})= 153-122 {\rm Jy} $. 

For each daily observation of Cygnus A we check the beam of the instrument 
 by fitting a two-dimensional Gaussian with an allowed base-label offset. 
In the frequency range of COSMO11, the Rayleigh-Jeans limit applies.
Therefore, the antenna temperature associated with a given flux 
from Cygnus A is
\beq
\label{eq-T}
T(\nu) = {{c^2} \over 2 k} {S_{\rm{Cyg}}(\nu) \over { \nu^2 \, \Omega}},
\eeq
where $ c $ is the vacuum speed of light, $ k $ is the 
Boltzmann's constant and 
$ \Omega = 
(\pi/4 \ln2 )\, \rm{ FWHM}^2 $ is the solid angle subtended by the beam. 

The effect of the band-pass filters is to yield a weighted 
antenna temperature for each observation of Cygnus A according to
\beq
T_{\,\rm{Cyg A}}^{\small C} = {\large\int}_{\nu_{min}}^{\nu_{max}}
\rho^{\small C}(\nu) 
T^{B}_{\,\rm{Cyg A}} (\nu) \, d\nu,
\eeq 
where the script $C$ stands for channel and $ T_{\,Cyg}^B $ is simply 
Eq.~\ref{eq-T} with $ S$ replaced by Eq.~\ref{eq-S} and 
where $ \rho^{\small C} (\nu) $ is the 
normalized linear spectral response of each filter. 
Assuming a circular FWHM of 1$^\circ$ the antenna temperature results
110.2 mK and 110.9 mK for channel 1 and
2 , respectively. For the fluxes, we get 137.9 Jy and 
138.3 Jy for channel 1 and 2, respectively. This shows that the differences 
due to non
identical band-pass filters are of order $\sim$ 0.5 \% only. In 
Sect.~\ref{sec-maps} we plot the distribution of daily root mean square
(RMS) in clean regions.

On the other hand, for a beam size of $1^{\circ}$, the polarization of the Cygnus A
region is less than 3-5 \%. We have not attempted to perform a detail 
study of the expected difference of the calibration for each of the two channels 
taking into account the COSMOSOMAS scanning strategy and data processing.
The reason is that, as explained later, the cumulative maps of each channel are recalibrated 
against unpolarized HII regions. Indeed, this results in less than 3 \% difference 
in the calibration of
both channels for the Cygnus region. As we will see, in this work we do not need to go to 
that precision in the data analysis, since, typical, statistical errors are of order 3-5\% also.

 We remark that the error associated with the  Baars et al. model is dominated
 by the rather large  uncertainty in the spectral indexes $ a $ and $ b$.
However, we have checked that this model predicts flux values for Cygnus A 
at WMAP frequencies within 10 \% of the WMAP observed fluxes. 
 
Finally, as an indication, for a good day, the  mean RMS of data in regions
placed at high Galactic latitude ($|b| > 50^{\circ}$) is
of order 1.1--1.5 mK, close to the theoretical sensitivity limit of the
receivers.

\subsection{Cumulative maps}
\label{sec-maps}
In this section, we present the first cumulative maps  obtained with  the 
COSMO11 experiment. Maps are pixelized in a rectangular RA and Dec grid. Temperatures for
each pixel are computed  as the weighted contribution of all the observations relevant to that
 pixel obtained through the whole observing campaign. The measurements associated to each pixel 
are weighted each day  with   the inverse of the variance of the data obtained that day
in a $3^\circ \times 3^{\circ}$ area centred in  the pixel under consideration. This
applies to all pixels  except those possibly affected by relatively bright  radio sources 
which are  treated in a different way. In these cases, the daily weight   is taken
as the inverse of the variance of the data  in high Galactic latitude regions where
no strong radio sources are known to contribute. Pixels within the WMAP ``Kp0'' mask 
(Bennet et al. 2003) are considered in this category.

\begin{figure}
\begin{center}
\includegraphics[height=6cm,width=16cm,angle=0]{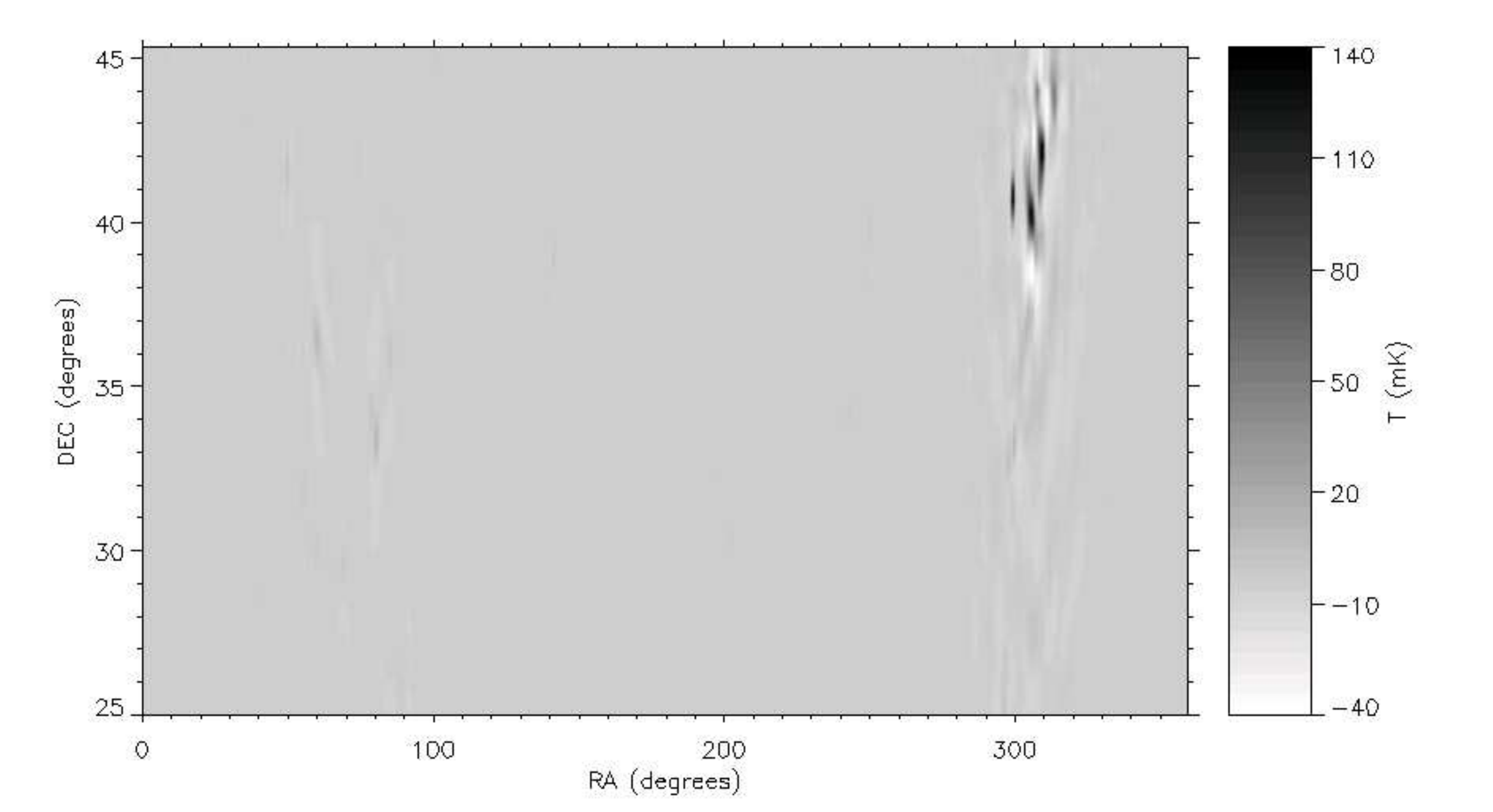}
\includegraphics[height=6cm,width=16cm,angle=0]{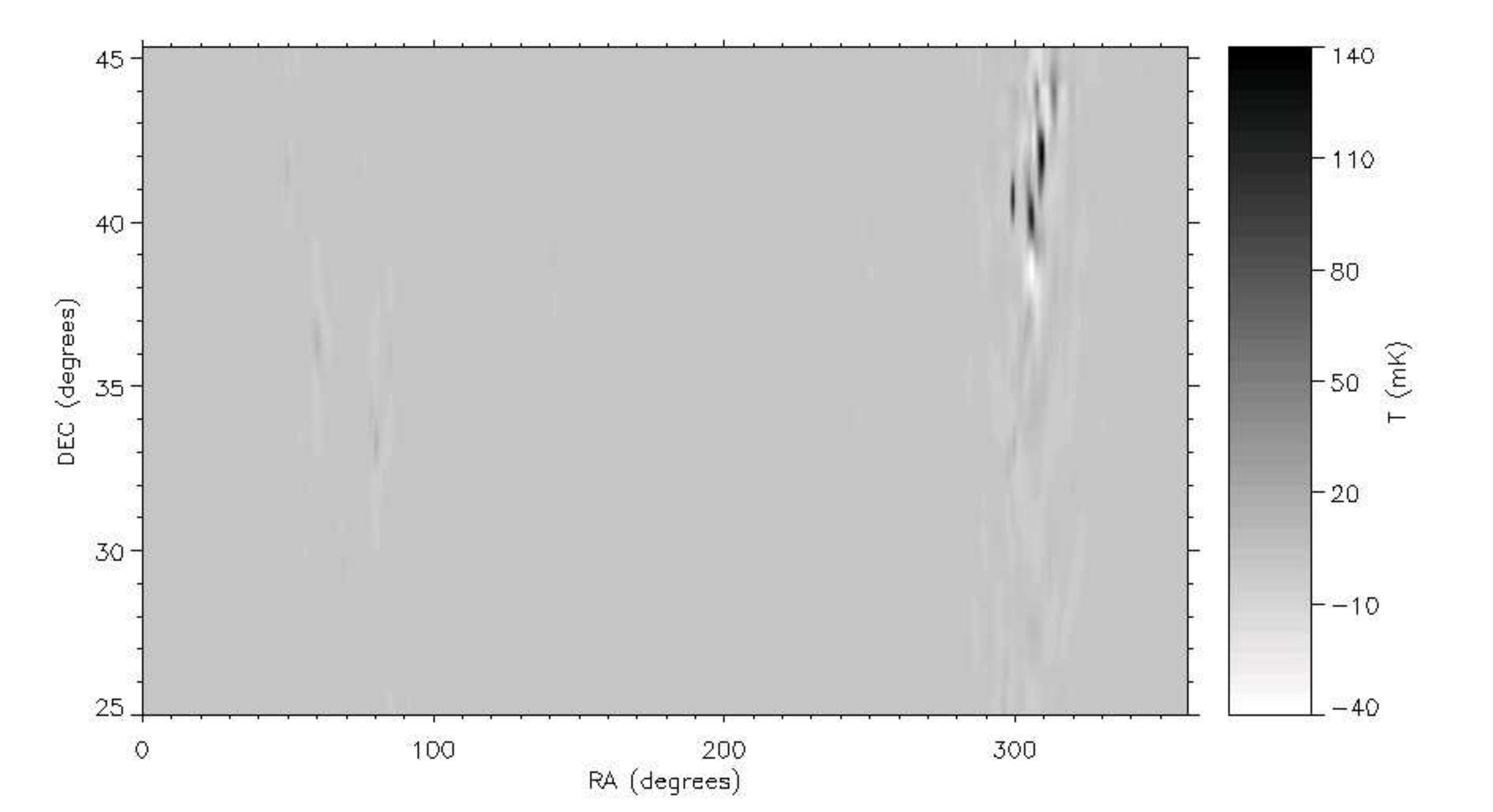}
\caption{Stacked map of COSMO11 channel 1 (top) and channel 2 (bottom) 
for the period November 2003 to June 2005.}
\label{fig-map11_nov03_jun05}
\end{center}
\end{figure}

 In Fig.~\ref{fig-map11_nov03_jun05} we present the final maps of 
COSMO11 obtained according to the data processing  described in Sect.~\ref{sec-day}, i.e., 
removing a $7^{\rm th}$ order Fourier series along each scan  and a $4^{\rm th}$ order 
one along day-night data to remove  atmospheric contamination to the highest possible level.
 As mentioned in the previous  section, the map for  channel 1 has the equivalent of 135 days 
of observations and  for channel 2 of 140 days. The angular resolution slightly changes along this map. 
The values measured for Cygnus A in these two maps, a point source 
when compared with our beam, are  listed 
in Tab.~\ref{fwhm11}.  In the final maps, for the region 
120$^{\circ} \le {\rm RA} \le 270^{\circ} $ and
25$^{\circ} \le {\rm DEC} \le 45^{\circ} $, the RMS per beam  (out of
the masked regions) is approximately $ 52 \mu K$  and $ 48  \mu K $ 
for channel 1 and 2, respectively.

\onecolumn
\twocolumn

\begin{figure}
\begin{center}
\includegraphics[height=5cm,width=8cm,angle=0]{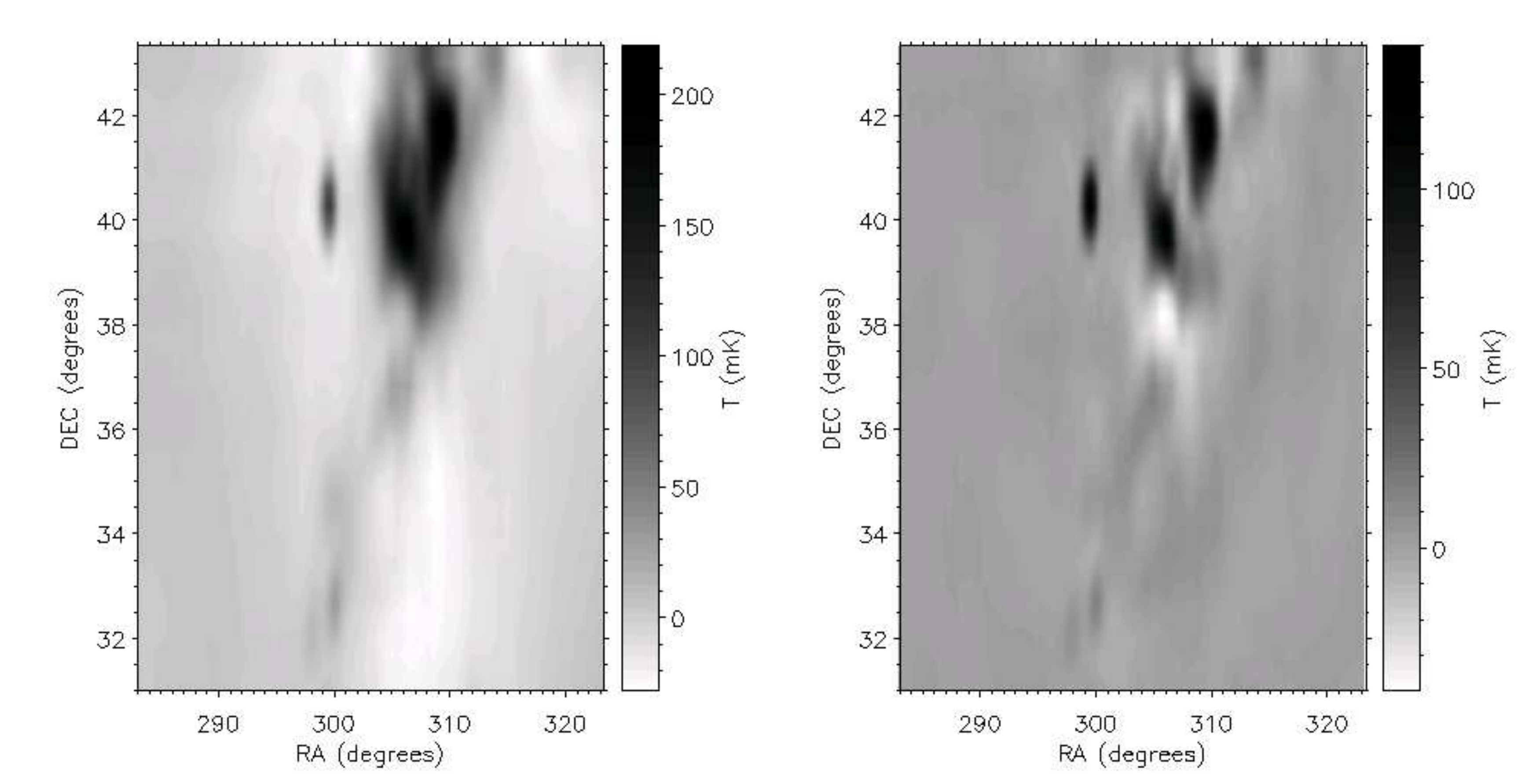}
\caption{Left: Galactic plane image in the cumulative map of channel
2 where a 1$^{\rm st}$ order Fourier series has been subtracted along
each scan. Right: Same, but with the present 7$^{\rm th} $ order
Fourier series atmospheric subtraction.}
\label{fig-galactic-center-1-3}
\end{center}
\end{figure}

 Finally, as mentioned in Sect.~\ref{sec-day}, the  COSMO11 data are not so severely affected
by the atmosphere as the COSMO15 data. This   may allow a correction of the atmospheric contamination using
 only low-order harmonics in the Fourier series and thus preserving information on larger angular  scales 
in the final map. In Fig.~\ref{fig-galactic-center-1-3} we show the cumulative map  obtained
in a  region of the   Galactic Plane following the same fitting procedure
described in Sect.~\ref{sec-day} but restricted to use only first order harmonics
in the Fourier series fit applied to each scan. The resulting maps (similar for both channels) are clearly
richer at large scales than the  maps constructed with higher order harmonic subtraction.
The full exploitation of the information provided by COSMO11 at large
 scales is left for a future  work. Here, we just  verified 
that the values given in Tab.~\ref{fwhm11} are compatible with
those measured in maps generated with  atmospheric removal using different number of harmonics
in the Fourier series. The result is rather independent on  the number of harmonics which
  is taken as an indication that the implemented source masking procedure 
 works properly for our main calibrator.

\begin{table}
\caption{FWHM in stacked maps for Cygnus A}
\vspace{0.3cm}
\begin{tabular}{c c c} 
 & $ {\rm FWHM}_{\rm RA} ({}^\circ) $  & $ {\rm FWHM}_{\rm DEC}
({}^\circ) $ \\
\hline
\hline
COSMO11 channel 1 &  0.80 $\pm$ 0.01 & 1.03 $\pm$ 0.01\\ 
\hline
COSMO11 channel 2 &  0.80 $\pm$ 0.01 &  1.01 $\pm$ 0.01\\ 
\hline
\end{tabular}
\label{fwhm11}
\end{table}

\begin{figure}
\begin{center}
\includegraphics[height=5cm,width=8cm,angle=0]{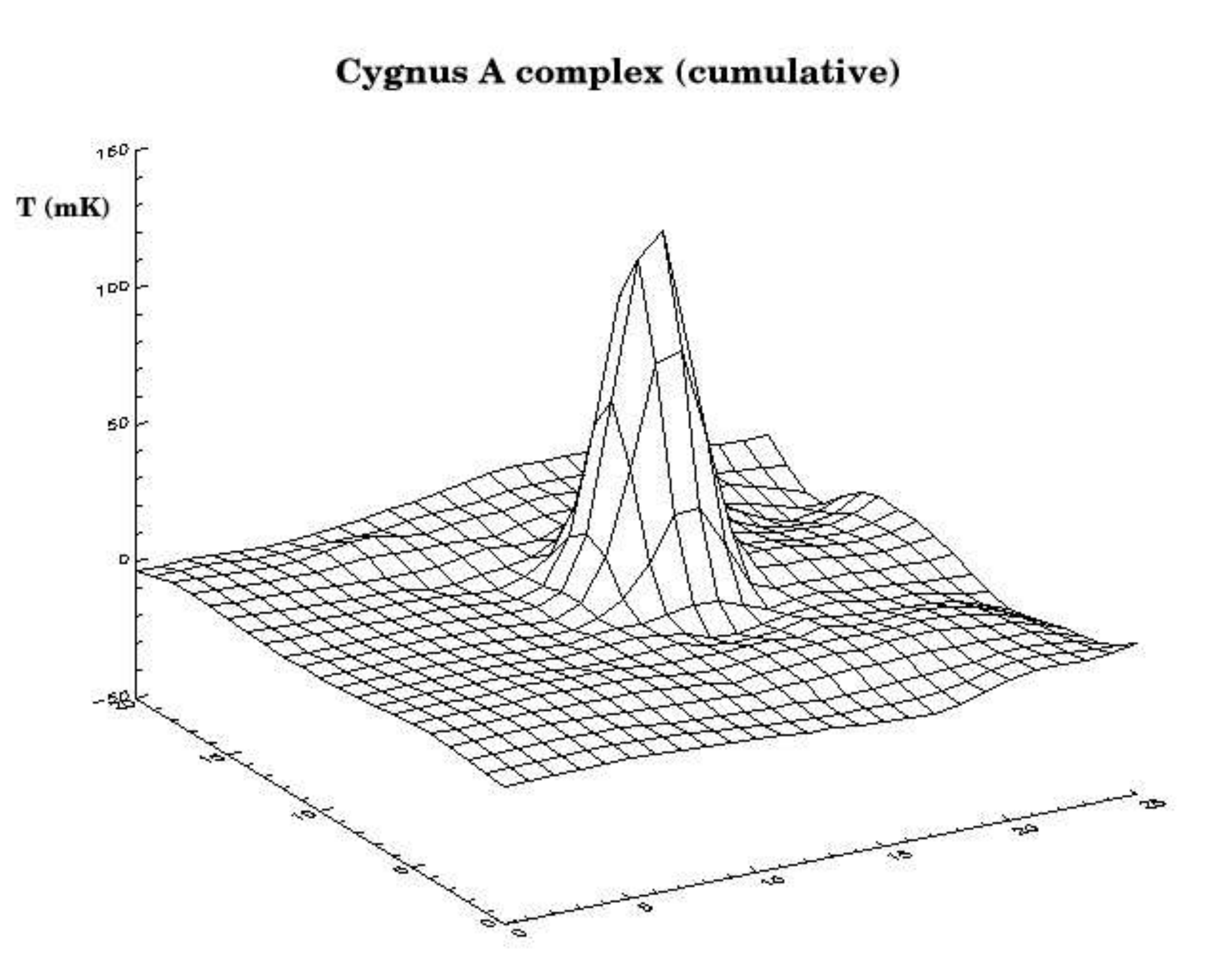}
\includegraphics[height=5cm,width=8cm,angle=0]{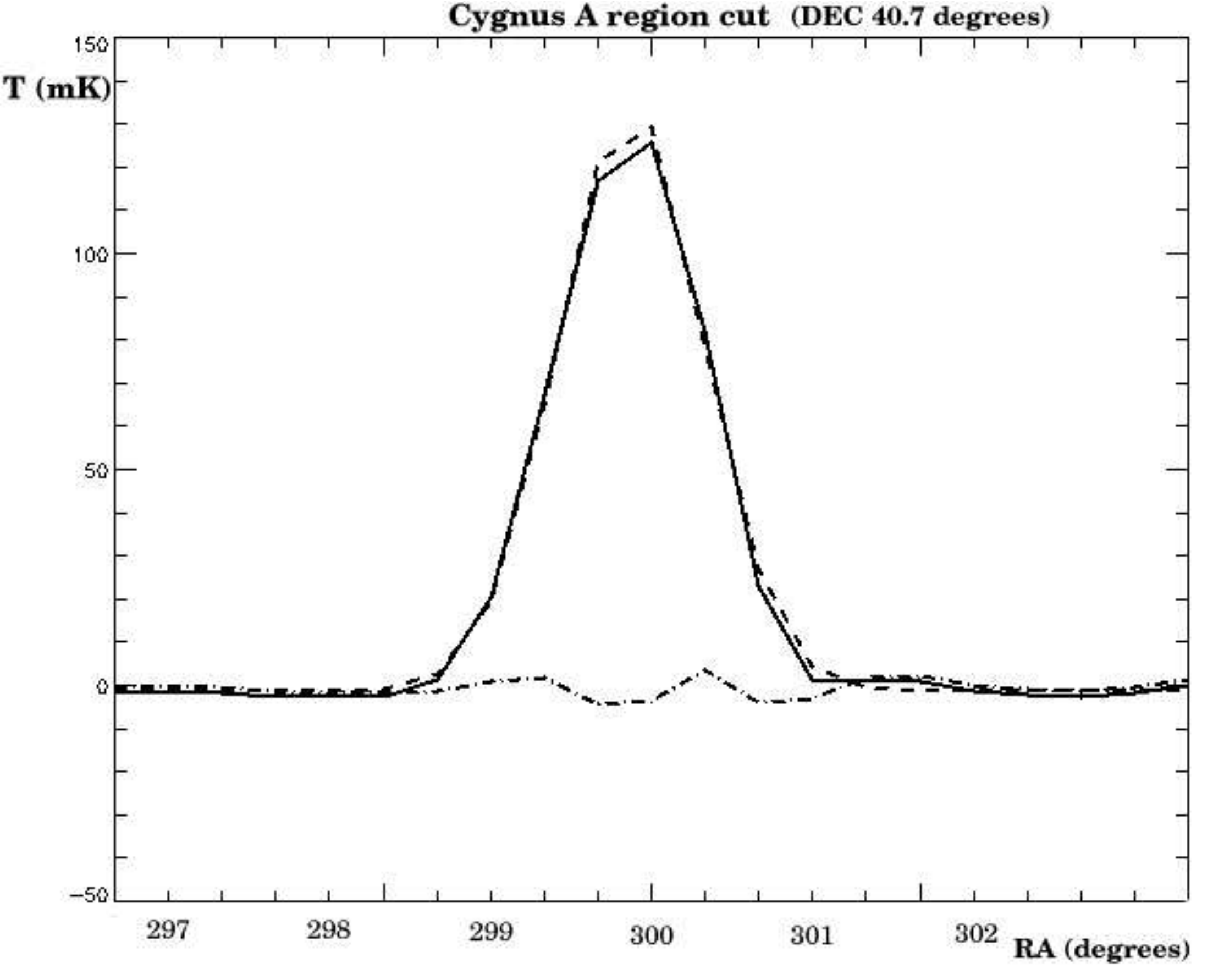}
\caption{Top: Image of the Cygnus A complex for the period November 2003
to June 2005. 
Bottom: Best 2-dimensional Gaussian fit to Cygnus A (dashed line). Observations (solid
line) and difference with the fit (dot-dashed line). Cuts are plotted at the declination
of the source.}
\label{}
\end{center}
\end{figure}

\subsection{Source recognition}
\label{sec-sources}
 A search for  radiosources was performed with SExtractor (Bertin \& Arnots 1996) 
on the map resulting from the combination of the two channels.  Some of the brightest point sources  
are listed  in Tab.~\ref{tab-sources11} with its most
likely identification.
The coordinates given in the table are those provided by  SExtractor and
were used to identify the sources in the K\"uhr \citep{kuhr}, 3C ({\tt http://www.mrao.cam.ac.uk/}) 
and Green Bank \citep{gb6} catalogues. 
We identify each radio source as the brightest   source ($> 1 \,$Jy) found in these
catalogues  within 15 arcminutes of the COSMOSOMAS pointing.
The 11 GHz fluxes  were obtained for these sources  as in Fern\'{a}ndez-Cerezo et al (2006) 
by fitting the theoretical point spread function to the COSMO11 data. The errors listed in the table 
include the calibration uncertainty. 
The WMAP catalogue was also searched and  22 GHz fluxes are listed whenever available. 
The sources not found in the WMAP catalogue are
either strong steep spectrum sources such as 3C123 or weaker flat spectrum
sources such as J2202+4216 which were too faint at the WMAP frequencies to be
detected.

Our results are generally consistent with independent observations of 
these radiosources at similar frequencies \citep{kuhr}, except for the
source 3C84 for which we give much lower fluxes, but it is well known that
the flux of this radiosource has progressively declined in the last few years. 
The  flux measured by COSMO11 is compatible
with the tendency delineated by the observations at the UMRAO (Univ. of Michigan Radio Astronomy 
observatory) in the past decade. Many of the brightest radio sources show large variability
and therefore the comparison of fluxes taken in different epochs can only provide  a qualitative
assessment of our data.

\begin{figure}
\begin{center}
\includegraphics[height=8cm,width=9.5cm,angle=0]{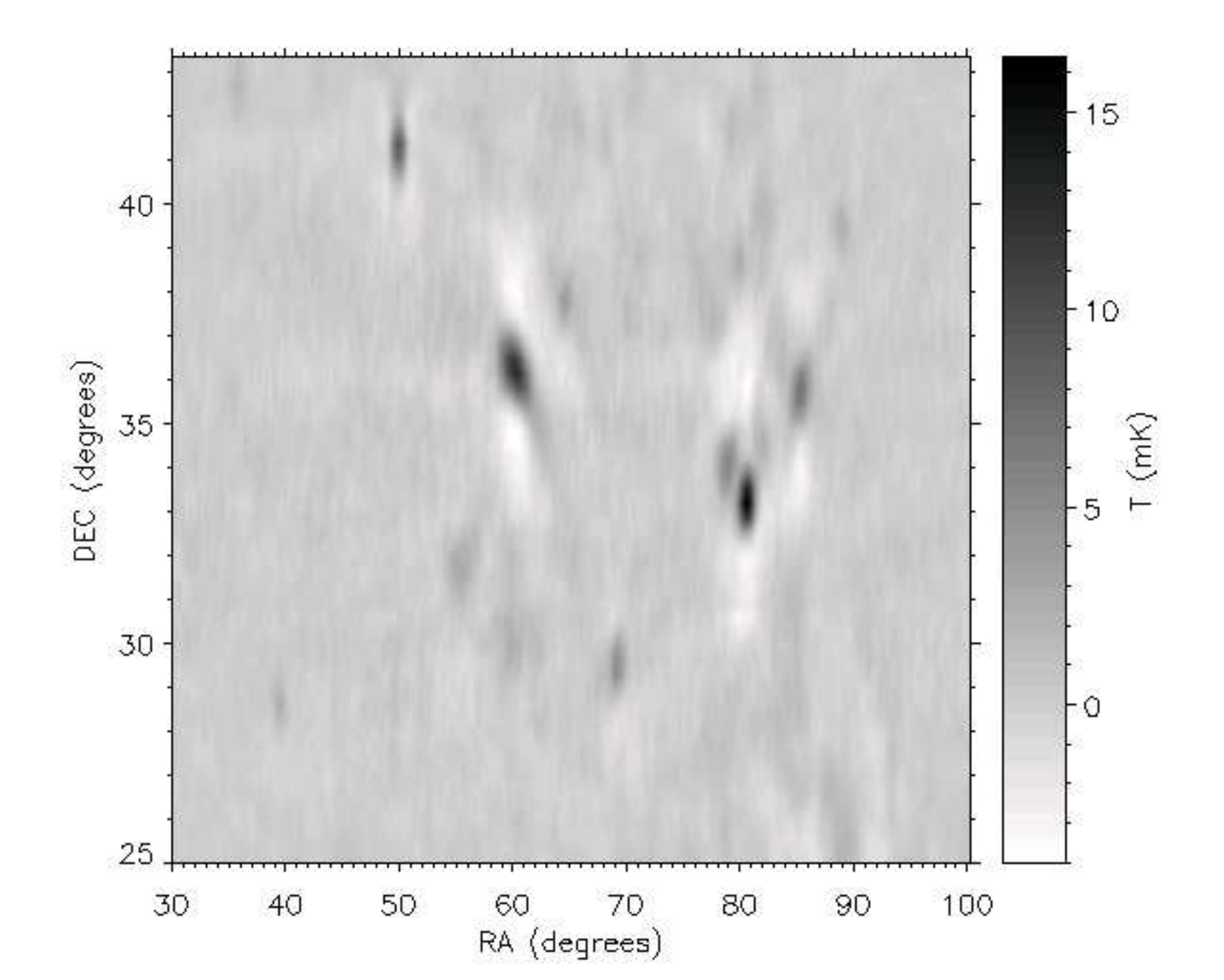}
\caption{A section of the final  map of  COSMO11 (channel 1) 
showing the detection of point-like sources towards the Galactic anticenter.}
\label{sources11}
\end{center}
\end{figure}

 We shall note that our  ability to detect point radiosources in   the COSMO11 data 
at high Galactic latitudes is effectively limited to sources with  fluxes above $\sim$ 1 Jy 
due to statistical noise and to the  ``confusion noise" associated to  an unresolved population of  
extragalactic radio sources. In Fig.~\ref{sources11}, a zoom of the
final map of COSMO11 shows the detection of several point-like sources near the Galactic 
plane in the anticenter direction.

\begin{table*}
\label{tab-sources11} 
\caption{A selection of the brightest sources detected in the COSMO11 channels and
the corresponding value in the WMAP K channel for comparison.}
\begin{tabular}[p]{rrrrrr} \hline 
Source     & R.A.   & Dec.  & Channel 1 flux (Jy) & Channel 2 flux(Jy) & WMAP K \\ \hline 
3C84       & 50.16  & 41.54 & 16.6 $\pm$ 1.7 & 17.8 $\pm$  1.8 & 11.1 \\ 
3C123      & 69.18  & 29.33 & 14.0 $\pm$ 1.4 & 13.7 $\pm$  1.4 &--- \\ 
0923+39    & 141.91 & 38.89 & 10.2 $\pm$ 1.0 & 10.8 $\pm$  1.1 &  6.8 \\ 
J0418+3801 & 64.64  & 38.08 & 9.5 $\pm$ 1.0 & 9.0 $\pm$  0.9 &--- \\ 
3C345      & 250.74 & 39.86 & 8.6 $\pm$ 0.9 & 9.2 $\pm$  1.0  &  8.0 \\ 
J0555+3948 & 89.17  & 39.79 & 3.9 $\pm$ 0.4 & 4.7 $\pm$  0.5 &  1.2 \\ 
3C465      & 354.73 & 26.65 & 3.2 $\pm$ 0.4 & 3.5 $\pm$  0.4 &--- \\ 
\end{tabular} 
\end{table*}

\subsection{Noise analysis}
\begin{figure}
\begin{center}
\includegraphics[height=8cm,width=8cm,angle=0]{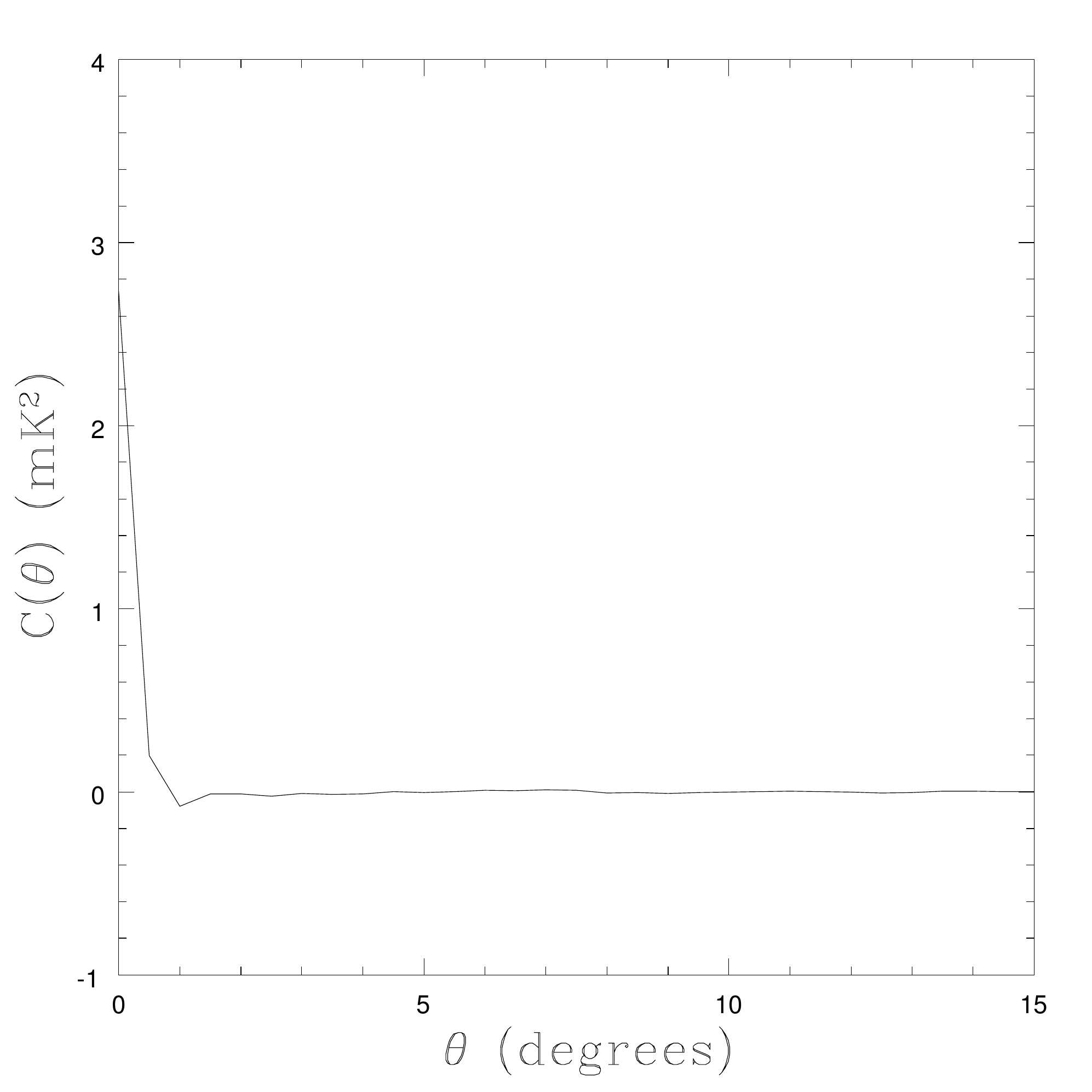}
\caption{Two-point correlation function for a single day in
COSMO11.}
\label{correlation-function11}
\end{center}
\end{figure}

In this section, we describe the main features of the  noise that
is still present after subtraction of the first Fourier terms in the daily data. 
In Fig.~\ref{correlation-function11} we plot the  two-point correlation function
for a representative day.  This is performed in a region RA=(115$^\circ$, 267$^\circ$), 
DEC=(25$^\circ$, 45$^\circ$)
that corresponds to high Galactic latitude and 
using the source 
mask described in Sect.~\ref{sec-correlations}. 

Some degree of correlation is  expected between channels  observing simultaneously the  same sky region. 
As  mean covariance of the data of the two channels,  we have obtained the average of the two-point 
correlation functions  computed for  each day whenever  simultaneous observations  in both channels
 were available (98 \% of the days).  The result is 0.40 mK, whereas  the typical  daily   RMS of the 
data  is  $2.25$ mK and $1.96$ mK  for channel 1 and 2, respectively. For comparison, the correlation
 between different days was found to be less than  1 \%  in good agreement with  expectations for the
 amplitude of the astronomical signals.

\subsubsection{Stacked maps}
 Finally, we have also considered the evolution of the noise per pixel when stacking
daily maps. We verified that the RMS of the data contributing to  each pixel decreased
 roughly as $1/\sqrt{N}$,
where $ N $ is the number of stacked daily maps. In the
final maps of channel 1 and 2,  the typical errors in  temperature (per pixel) are of order 130-140 $\mu$K.

\section{Data analysis}
\label{sec-correlations}
In this section, we study the degree of correlation of the previous two COSMO11 
stacked maps with  multifrequency datasets available in the literature for the same sky region. 
The goal is to establish  the contribution of CMB, synchrotron, brehmstralung, extragalactic radiosources  and 
anomalous microwave emission at the frequency and angular scales of our experiment. The set of maps to be correlated with COSMO11 are:
(i) the 3rd year maps for each of the five  channels of the WMAP mission 
\citep{hinshaw07}\footnote{These maps are given in the HEALPix pixelisation scheme\citep{healpix}.};
(ii) the extragalactic radiosource emission maps constructed from the NVSS catalogue;
(iii) the 408 MHz map provided by \citep{haslam82};
(iv) the  $ {\rm H}_{\alpha} $ map provided by~\citep{finkbeiner03};
and (v) the DIRBE maps at 100 $ \mu $m (DIRBE08) and  240 $\mu$m (DIRBE10), and 
a HEALPix version of the dust map 
provided by \citep{SFD} from LAMBDA~\footnote{http://lambda.gsfc.nasa.gov/.} at 100 $\mu$m,
($ \Lambda_{100} $ in the sequel).

 The method used for the computation of the correlation
between maps is described in detail in de Oliveira-Costa et al. (2004) and
Fern\'{a}ndez-Cerezo et~al. (2006)
Correlation coefficients can be computed by comparison with just one  template map
or simultaneously with several maps. The multi-correlation option  is preferred when the templates present
some degree of correlation themselves, as it is the case of those describing Galactic emission processes.
In this work, all the correlation results except those  corresponding to CMB come from a multi
correlation analysis (unless otherwise stated)  were the templates considered  are: i) the V channel of the WMAP mission 
for CMB,  
ii) the NVSS map, iii) the 408 MHz map, iv) the H$_{\alpha}$ map and v) any 
of the three dust maps mentioned before.

\begin{figure}
\begin{center}
\includegraphics[width=6cm,angle=0]{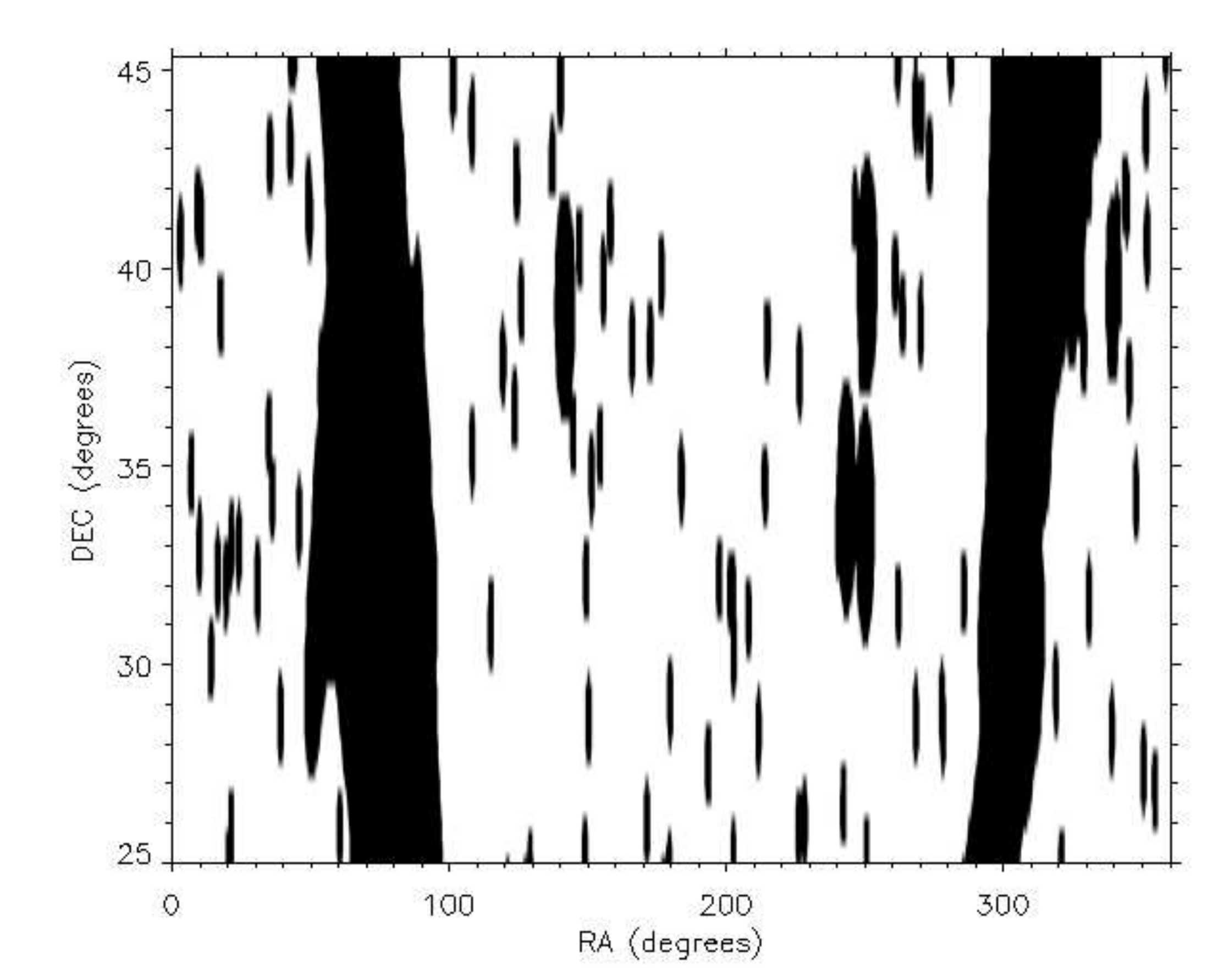}
\caption{Mask used for the  cross-correlation studies, see text for details.}
\label{fig-ps_mask}
\end{center}
\end{figure}

The correlation study considers data between
$ 25.0^\circ \le {\rm DEC} \le 44.7^\circ $  as allowed by the
mask plotted  in Fig.~\ref{fig-ps_mask}.  This is an extended version 
of the ``Kp0'' mask used by WMAP where  any adjacent pixel  
 was also discarded for the correlation. Additionally, we masked 
some regions at high galactic latitude with strong radiosources to minimize any possible 
residual effects  of these sources after the adopted atmospheric filtering. The regions masked are (RA;DEC): 
(141.7$^\circ \pm 3.7^{\circ}, 38.7^{\circ}\pm 2.7^{\circ}$),
($243.0^{\circ}\pm 3.7^{\circ},34.3^{\circ}\pm 3.3^{\circ}$),
($250.7^{\circ} \pm 3.7^{\circ}, 33.5^{\circ} \pm 3.3^{\circ}$),
($250.7^{\circ} \pm 3.7^{\circ}, 39.7^{\circ} \pm 2.7^{\circ}$).

Several other masking criteria were also considered in 
the correlation analysis:  excluding all additional  sources for which a direct extrapolation of 
the fluxes in the  NVSS and GB6 catalogues yielded values  higher than 1 Jy at 11 GHz; 
 extending the Kp0 by two  or three pixels around each masked source; and finally,  excluding any  pixels
 with temperatures deviating several times the typical error per pixel of the map. The correlation 
results obtained with the various masks agreed within the statistical error bars of the method.

 In the following sections we deal with 1$\sigma$ statistical errors estimated directly
from the correlation method,  but other sources of error should also  be considered. Absolute calibration
errors amount  up to  10 \% and  chance alignment of regions with
 emission caused by physically unrelated processes also contribute to the final uncertainties.
To determine the role of chance alignment, we have computed the correlation values  and their  dispersion
 for  arbitrary rotations of the templates around an axis perpendicular to the 
Galactic plane. 
The dispersions of the
correlation values  were found  usually 1--2 times higher than the statistical error given by the correlation
 method.  Only those correlation values larger than 2 times the statiscal error will be considered 
a reliable result.

The COSMO11 maps are convolved to a final angular resolution of 1.12$^\circ \times $1.12$^\circ$ to set on  equal foot  the COSMO11 and COSMO15 data, so both experiments can be compared directly. 
We have not considered the different polarization of the two COSMO11 channels  in producing the 
set of templates. This may lead to systematic differences in the correlation results obtained for each channel with  the same templates. In general, these differences will not be larger than a few percent of the amplitude of the correlated signals, i.e. smaller than  the statistical error of the correlation method. 

 With the adopted  atmospheric filtering, the  window function describing the data of the COSMO11 radiometer
 is analogous to that published  by Fernández-Cerezo et al. (2006) for COSMO15. We remark that the
  filtering applied to the COSMOSOMAS data during the reduction process is also 
applied to any other map used in the correlation analysis. This  erases information on the largest
 scales, so the results need to be understood as the correlation level at the
remaining scales, which essentially range from $\sim 1^{\circ}$ to $5^{\circ}$.
Finally,  here we have reprocessed the COSMO15  data  obtained by Fern\'andez-Cerezo et al. (2006),
according to the procedures  explained in Sect ~\ref{sec-data} which are slightly different to those
used in the former work.  The  cross-correlation  results obtained 
for the COSMO15 data are also  slightly different.

\subsection{Correlations with WMAP data}
\label{sec-correlationswmap}
In Tab.~\ref{tbl-CMB} we list the correlation values obtained for the COSMOSOMAS (labelled as C11$_1$ and C11$_2$ for the two COSMO11 channels, and C13, C15 and C16 for the three  COSMO15 channels)  and the WMAP channels 
for three different galactic latitude cuts. First, we note that the correlations among the highest
frequency channels of WMAP indicate  the level of the CMB signal left by the window
function associated to the adopted COSMOSOMAS data processing scheme.  The correlations between the Q,  V and W channels 
show a detection of CMB with amplitude 26-26.5 $\mu$K at the various  Galactic latitude cuts. This is in agreement
with the expected CMB fluctuation after removal of the Fourier terms described in Sect.~\ref{sec-day}. The autocorrelation values for these   channels are $\sim$0.5 $\mu$K higher as expected for the noise contribution.

\begin{figure}
\begin{center}
\includegraphics[height=8cm,width=8cm,angle=0]{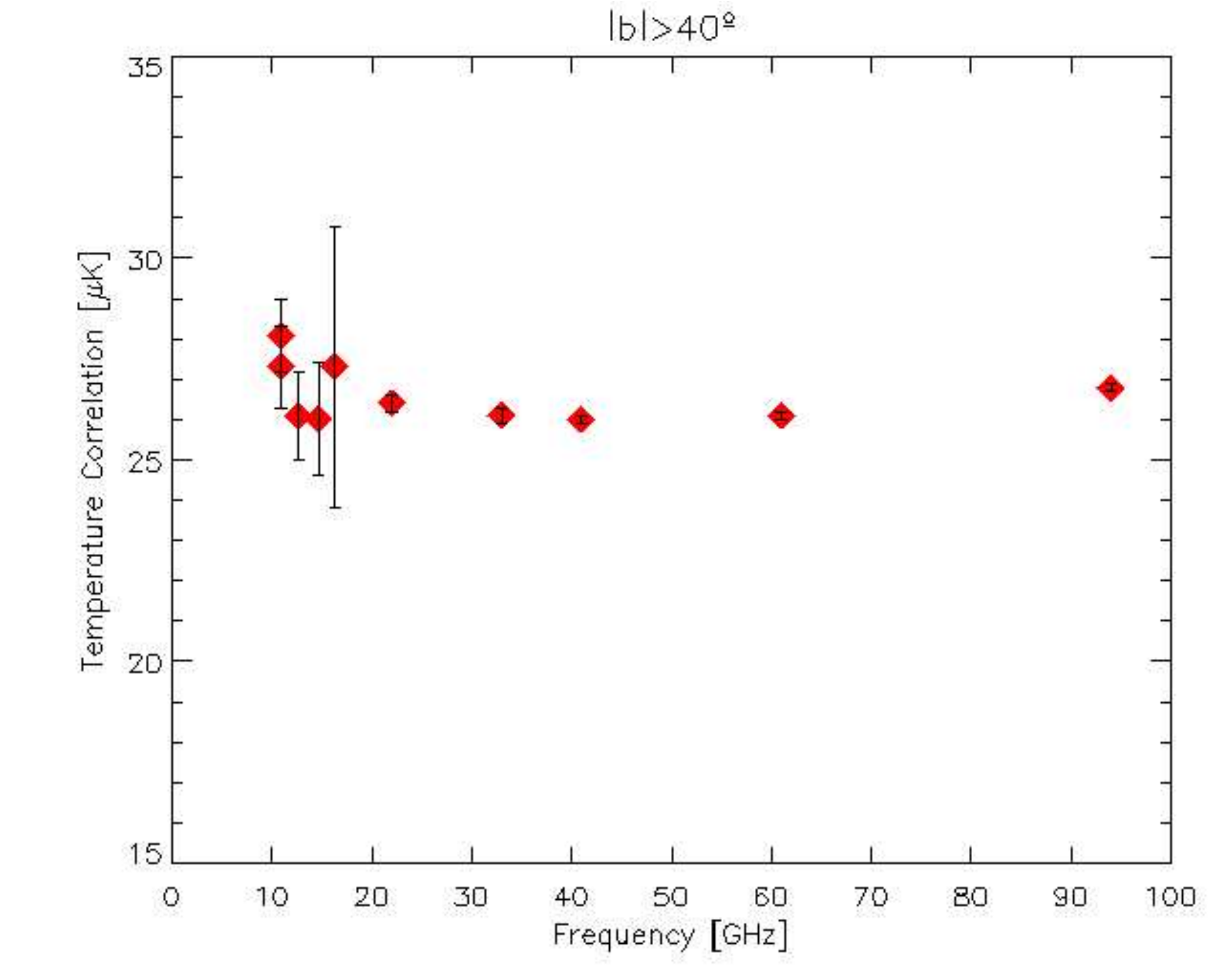}
\caption{Correlation of COSMOSOMAS and WMAP channels with
the V channel of WMAP. Error bars are 1 $\sigma$ statistical errors.}
\label{fig-cmb}
\end{center}
\end{figure}

Approximately the  same level of correlation (26-29 $\mu$K) is found between the five  COSMOSOMAS channels, 
and the high frequency  WMAP channels (see Fig.~\ref{fig-cmb}). The CMB signal is detected in both  COSMO11
 channels with a 
high significance ($>$ 25$\sigma$).  CMB polarization is expected to cause  differences between
 the correlation values obtained for each   of the  two COSMO11 channels smaller than the statistical
uncertainties.  Therefore no attempt was  made to measure this polarization effect in our 
data. When we consider regions at low Galactic latitude, as e.g. $|b|>30^{\circ}$,
the amplitude of the CMB correlated signal  slightly decreases, this is an artifact  due
 to  the strong influence of  the Galactic plane signal which  affects
 the processing of COSMOSOMAS data.

 The amplitude of the correlated signal between COSMO11 and the low-frequency  WMAP channels 
increases to $\sim$43 and 35 $\mu$K, 
for the K and K$_a$ channels ($|b|>40^{\circ}$), respectively.
 This  is attributed to signals in common other than CMB. At high galactic latitude and considering 
the angular scales relevant to the present analysis,  the expected major foreground contamination in common
 with these WMAP channels are unresolved  extragalactic radiosources as we will see  in the next sections.

\begin{table*}
\caption{Values of the temperature correlation in $ \mu $K for the
COSMO11 cumulative map with respect to the maps of the five WMAP
channels for different cuts in Galactic
latitude, $b$. Errors are 1 $\sigma$.}
\label{tbl-CMB}
{\tiny
\begin{center}
\begin{tabular}{lcccccccccc}
\hline
Template & ${\rm C11}_1$& ${\rm C11}_2$& C13& C15& C16& WMAP\_K& WMAP\_Ka& WMAP\_Q& WMAP\_V& WMAP\_W \\
\hline
\hline
\multicolumn{11}{c}{$|b|$$>30^\circ$}  \\
\hline
WMAP\_K & $\mathbf{42.5 \pm 0.9}$ & $\mathbf{43.8 \pm 0.8}$ & $\mathbf{32.1 \pm 1.0}$ & $\mathbf{30.9 \pm 1.2}$ & $\mathbf{31.9 \pm 3.0}$ & $\mathbf{29.3 \pm 0.1}$ & $\mathbf{26.0 \pm 0.1}$ & $\mathbf{25.3 \pm 0.1}$ & $\mathbf{24.7 \pm 0.1}$ & $\mathbf{24.4 \pm 0.1}$ \\
WMAP\_Ka & $\mathbf{33.0 \pm 0.9}$ & $\mathbf{34.0 \pm 0.8}$ & $\mathbf{27.3 \pm 1.0}$ & $\mathbf{26.3 \pm 1.2}$ & $\mathbf{28.4 \pm 3.0}$ & $\mathbf{28.0 \pm 0.1}$ & $\mathbf{27.4 \pm 0.1}$ & $\mathbf{26.5 \pm 0.1}$ & $\mathbf{26.2 \pm 0.1}$ & $\mathbf{26.0 \pm 0.1}$ \\
WMAP\_Q & $\mathbf{29.9 \pm 0.9}$ & $\mathbf{31.1 \pm 0.8}$ & $\mathbf{26.1 \pm 1.0}$ & $\mathbf{25.5 \pm 1.2}$ & $\mathbf{27.7 \pm 3.0}$ & $\mathbf{27.5 \pm 0.1}$ & $\mathbf{26.8 \pm 0.1}$ & $\mathbf{27.0 \pm 0.1}$ & $\mathbf{26.4 \pm 0.1}$ & $\mathbf{26.2 \pm 0.1}$ \\
WMAP\_V & $\mathbf{27.0 \pm 0.9}$ & $\mathbf{27.7 \pm 0.8}$ & $\mathbf{23.9 \pm 1.0}$ & $\mathbf{23.9 \pm 1.2}$ & $\mathbf{25.0 \pm 3.0}$ & $\mathbf{26.9 \pm 0.1}$ & $\mathbf{26.5 \pm 0.1}$ & $\mathbf{26.4 \pm 0.1}$ & $\mathbf{26.9 \pm 0.1}$ & $\mathbf{26.2 \pm 0.1}$ \\
WMAP\_W & $\mathbf{25.1 \pm 0.9}$ & $\mathbf{25.7 \pm 0.8}$ & $\mathbf{23.0 \pm 1.0}$ & $\mathbf{23.5 \pm 1.2}$ & $\mathbf{24.9 \pm 3.0}$ & $\mathbf{26.5 \pm 0.1}$ & $\mathbf{26.3 \pm 0.1}$ & $\mathbf{26.2 \pm 0.1}$ & $\mathbf{26.2 \pm 0.1}$ & $\mathbf{27.0 \pm 0.1}$ \\
\hline
\multicolumn{11}{c}{$|b|$$>40^\circ$}  \\
\hline
WMAP\_K & $\mathbf{42.7 \pm 1.0}$ & $\mathbf{44.1 \pm 0.9}$ & $\mathbf{34.2 \pm 1.1}$ & $\mathbf{32.8 \pm 1.4}$ & $\mathbf{33.3 \pm 3.5}$ & $\mathbf{29.1 \pm 0.2}$ & $\mathbf{25.8 \pm 0.2}$ & $\mathbf{25.2 \pm 0.1}$ & $\mathbf{24.7 \pm 0.1}$ & $\mathbf{24.4 \pm 0.1}$ \\
WMAP\_Ka & $\mathbf{34.4 \pm 1.0}$ & $\mathbf{35.6 \pm 0.9}$ & $\mathbf{29.9 \pm 1.1}$ & $\mathbf{28.3 \pm 1.4}$ & $\mathbf{29.9 \pm 3.5}$ & $\mathbf{27.7 \pm 0.2}$ & $\mathbf{27.1 \pm 0.2}$ & $\mathbf{26.2 \pm 0.1}$ & $\mathbf{26.0 \pm 0.1}$ & $\mathbf{25.8 \pm 0.1}$ \\
WMAP\_Q & $\mathbf{31.7 \pm 1.0}$ & $\mathbf{33.1 \pm 0.9}$ & $\mathbf{28.8 \pm 1.1}$ & $\mathbf{28.0 \pm 1.4}$ & $\mathbf{29.6 \pm 3.5}$ & $\mathbf{27.4 \pm 0.2}$ & $\mathbf{26.5 \pm 0.2}$ & $\mathbf{26.7 \pm 0.1}$ & $\mathbf{26.3 \pm 0.1}$ & $\mathbf{26.1 \pm 0.1}$ \\
WMAP\_V & $\mathbf{29.0 \pm 1.0}$ & $\mathbf{29.7 \pm 0.9}$ & $\mathbf{26.9 \pm 1.1}$ & $\mathbf{26.5 \pm 1.4}$ & $\mathbf{27.2 \pm 3.5}$ & $\mathbf{26.8 \pm 0.2}$ & $\mathbf{26.3 \pm 0.2}$ & $\mathbf{26.2 \pm 0.1}$ & $\mathbf{26.8 \pm 0.1}$ & $\mathbf{26.1 \pm 0.1}$ \\
WMAP\_W & $\mathbf{27.3 \pm 1.0}$ & $\mathbf{28.1 \pm 0.9}$ & $\mathbf{26.1 \pm 1.1}$ & $\mathbf{26.0 \pm 1.4}$ & $\mathbf{27.3 \pm 3.5}$ & $\mathbf{26.4 \pm 0.2}$ & $\mathbf{26.1 \pm 0.2}$ & $\mathbf{26.0 \pm 0.1}$ & $\mathbf{26.1 \pm 0.1}$ & $\mathbf{26.8 \pm 0.1}$ \\
\hline
\multicolumn{11}{c}{$|b|$$>50^\circ$}  \\
\hline
WMAP\_K & $\mathbf{41.1 \pm 1.1}$ & $\mathbf{41.3 \pm 1.0}$ & $\mathbf{33.6 \pm 1.3}$ & $\mathbf{31.0 \pm 1.6}$ & $\mathbf{33.5 \pm 3.9}$ & $\mathbf{29.3 \pm 0.2}$ & $\mathbf{26.3 \pm 0.2}$ & $\mathbf{25.8 \pm 0.1}$ & $\mathbf{25.3 \pm 0.1}$ & $\mathbf{25.0 \pm 0.1}$ \\
WMAP\_Ka & $\mathbf{34.2 \pm 1.1}$ & $\mathbf{34.3 \pm 1.0}$ & $\mathbf{29.9 \pm 1.3}$ & $\mathbf{26.9 \pm 1.6}$ & $\mathbf{30.6 \pm 3.9}$ & $\mathbf{28.1 \pm 0.2}$ & $\mathbf{27.4 \pm 0.2}$ & $\mathbf{26.7 \pm 0.1}$ & $\mathbf{26.5 \pm 0.1}$ & $\mathbf{26.2 \pm 0.1}$ \\
WMAP\_Q & $\mathbf{31.6 \pm 1.1}$ & $\mathbf{32.1 \pm 1.0}$ & $\mathbf{28.9 \pm 1.3}$ & $\mathbf{26.7 \pm 1.6}$ & $\mathbf{30.7 \pm 3.9}$ & $\mathbf{27.8 \pm 0.2}$ & $\mathbf{26.9 \pm 0.2}$ & $\mathbf{27.1 \pm 0.1}$ & $\mathbf{26.6 \pm 0.1}$ & $\mathbf{26.4 \pm 0.1}$ \\
WMAP\_V & $\mathbf{29.1 \pm 1.1}$ & $\mathbf{29.0 \pm 1.0}$ & $\mathbf{27.2 \pm 1.3}$ & $\mathbf{25.3 \pm 1.6}$ & $\mathbf{28.4 \pm 3.9}$ & $\mathbf{27.3 \pm 0.2}$ & $\mathbf{26.7 \pm 0.2}$ & $\mathbf{26.7 \pm 0.1}$ & $\mathbf{27.1 \pm 0.1}$ & $\mathbf{26.4 \pm 0.1}$ \\
WMAP\_W & $\mathbf{28.0 \pm 1.1}$ & $\mathbf{27.9 \pm 1.0}$ & $\mathbf{26.9 \pm 1.3}$ & $\mathbf{25.7 \pm 1.6}$ & $\mathbf{29.0 \pm 3.9}$ & $\mathbf{26.9 \pm 0.2}$ & $\mathbf{26.5 \pm 0.2}$ & $\mathbf{26.5 \pm 0.1}$ & $\mathbf{26.5 \pm 0.1}$ & $\mathbf{27.1 \pm 0.1}$ \\
\hline
\hline
\end{tabular}
\end{center}
}

\end{table*}

\subsection{Correlations with unresolved radio sources}
\label{sec-urrs}
\begin{figure}
\begin{center}
\includegraphics[height=8cm,width=8cm,angle=0]{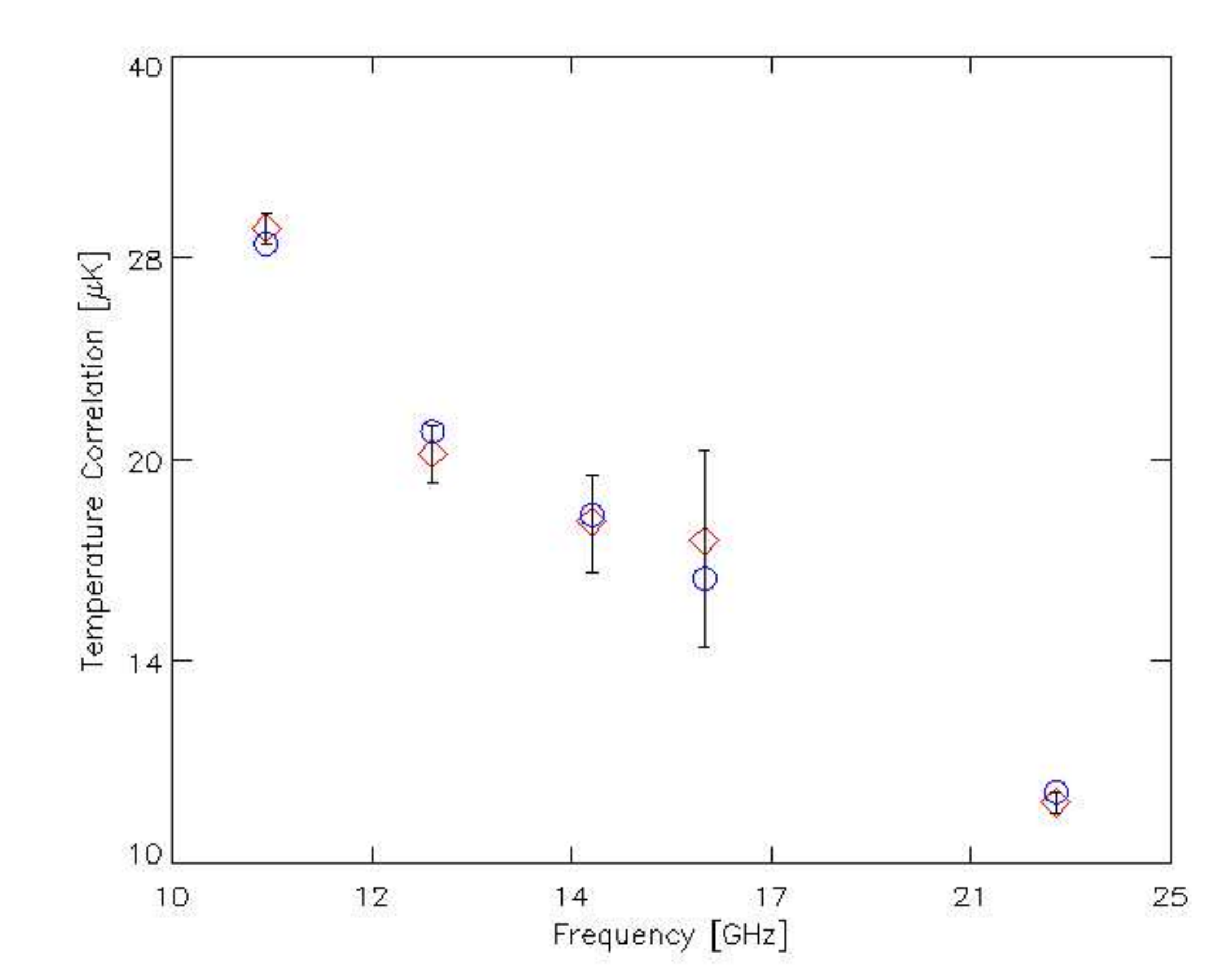}
\caption{Predicted (circles) and observed (diamonds) 
contribution from unresolved radio sources, after taking into account
COSMOSOMAS observing strategy and data processing. See text, and
Tab.~\ref{tab-excess} for details.}
\label{fig-excess_models}
\end{center}
\end{figure}

In order to determine the level of the contribution of unresolved radio sources in the COSMOSOMAS data
 we proceed as  in Fern\'{a}ndez-Cerezo et~al. (2006).
Essentially, for each  galactic latitude and  at each frequency
the  correlation values  of COSMOSOMAS with the WMAP K map (values of Tab.~\ref{tbl-CMB}) are corrected
  from the CMB contribution (obtained as  mentioned above), as well as from 
the  corresponding correlation with the Galactic synchrotron (408 MHz map), free-free ( H$\alpha$ template) and dust (DIRBE08)  (see next sections for details on  these correlations and their  statistical errors). 
The correction consists in a subtraction of  each of the previous temperature correlation values in quadrature. 
We obtain the  results listed in Tab.~\ref{tab-excess} and plotted  in Fig.~\ref{fig-excess_models}.
The  data point at 23 GHz in Fig.~\ref{fig-excess_models} 
 is obtained correcting the autocorrelation value of the WMAP K for the  contribution of the  CMB
 and the other foregrounds. It is clear the
 existence of  an additional  signal in common between the  COSMOSOMAS data and WMAP K band whose
amplitude is independent of galactic  latitude and  decreases with increasing frequency. 
 At the frequencies of COSMOSOMAS and at the angular scales under consideration,
this additional signal turns out to be the second most relevant contribution (after CMB) to the temperature fluctuations  at high Galactic latitudes.

\begin{table*}
\caption{Temperature cross correlation excess between the COSMOSOMAS and WMAP K map
after contributions from CMB and Galactic emission are accounted, see text for details.
All the maps are processed according to COSMOSOMAS strategy.
Errors are 1 $\sigma$. The error bars are computed taking into account the error
bars of each correlation and the cross correlation between templates.}
\label{tab-excess}
{\tiny
\begin{center}
\begin{tabular}{lcccccc}
\hline
Template & ${\rm C11}_1$& ${\rm C11}_2$& C13& C15& C16& WMAP\_K \\
\hline
\hline
\multicolumn{7}{c}{$|b|$$>30^\circ$}  \\
\hline
Excess of signal & $30.0 \pm 1.9$ & $30.6 \pm 1.7$ & $20.1 \pm 2.0$ & $18.4 \pm 2.5$ & $18.1 \pm 4.9$ & $11.3 \pm 0.4$ \\
\hline
\multicolumn{7}{c}{$|b|$$>40^\circ$}  \\
\hline
Excess of signal & $30.2 \pm 1.9$ & $31.0 \pm 1.8$ & $21.0 \pm 2.0$ & $18.7 \pm 2.8$ & $18.2 \pm 5.1$ & $11.1 \pm 0.4$ \\
\hline
\multicolumn{7}{c}{$|b|$$>50^\circ$}  \\
\hline
Excess of signal & $28.4 \pm 2.0$ & $28.8 \pm 1.8$ & $19.5 \pm 2.2$ & $16.9 \pm 2.9$ & $16.0 \pm 4.9$ & $10.7 \pm 0.5$ \\
\hline
\hline
\end{tabular}
\end{center}
}

\end{table*}

In order to interpret these results we use the models by de Zotti et al. (2005)
and  estimate the expected contribution of unresolved radio sources. 
Assuming a detection limit of  1 Jy for resolved radiosources (masked in our correlation
analysis) and considering our window function  these models predict
 at 11 GHz a total contribution of unresolved radiosources of  order 29.0 $\mu $K
(G\'onz\'alez-Nuevo private communication)
For the COSMO15 frequencies the models predict contributions of order
21.0 $\mu$K, 18.2 $\mu$K, 16.3 $\mu$K for C13, C15, and  C17 channels, respectively. 
Finally, for the WMAP K channel, the  predicted  contribution is 
of order 11.3 $\mu$K. The excess temperature correlation values in Tab.~\ref{tab-excess}  agree remarkably well
   with these model predictions. Both are compared  in  Fig.~\ref{fig-excess_models}.

\subsection{Correlations with 408 MHz and 1420 MHz data}
\label{sec-408MHz}
\begin{figure}
\begin{center}
\includegraphics[height=8cm,width=8cm,angle=0]{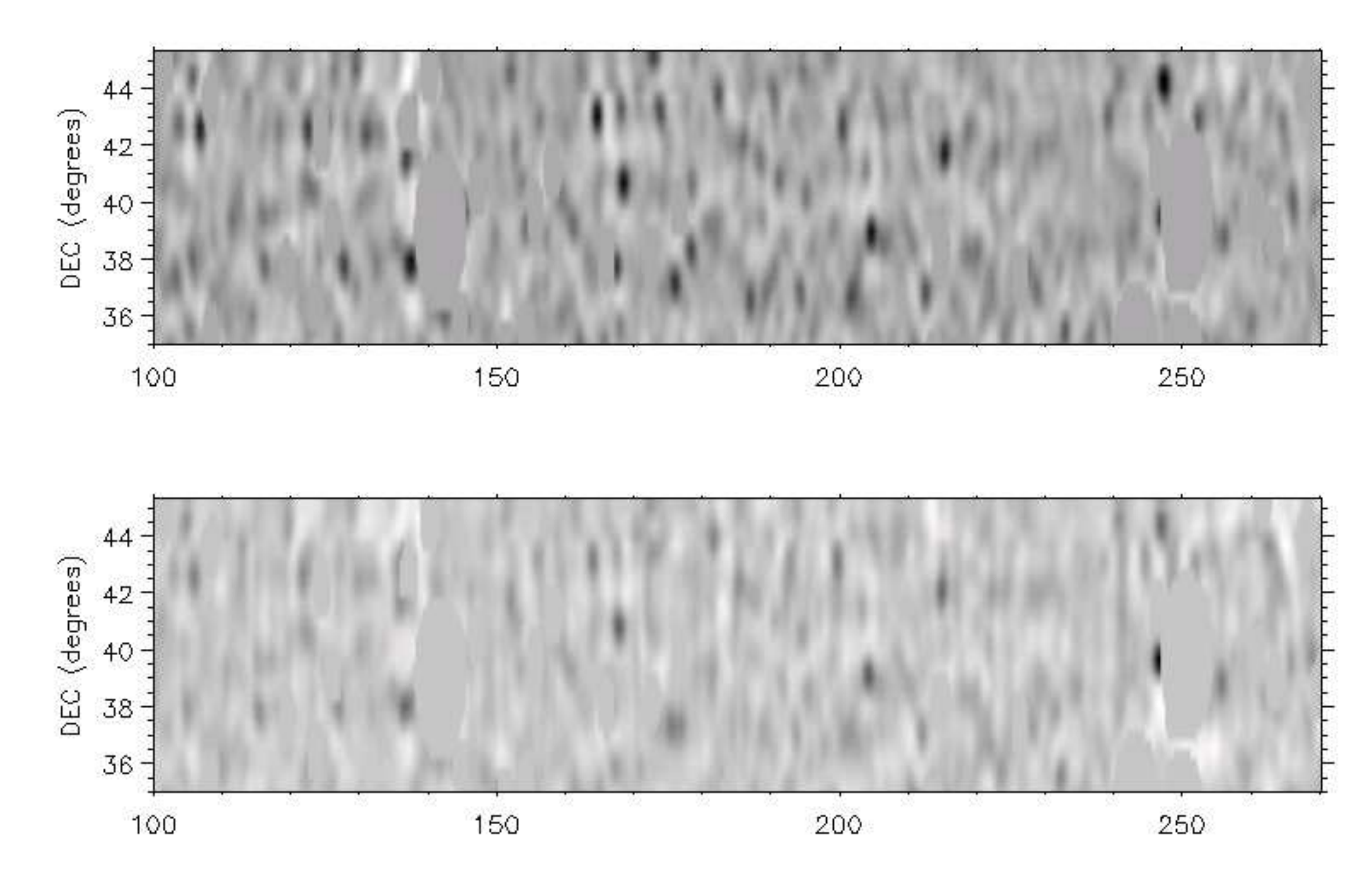}
\caption{A visual comparison between the NVSS map (top) and the 
408 MHz map (bottom) when processed according
to COSMOSOMAS strategy and degraded to a common 1 degree angular 
resolution.}
\label{fig-nvss-has}
\end{center}
\end{figure}

In the upper section of Tab.~\ref{tab-has} 
we list the correlation values between COSMOSOMAS,WMAP and the 408 MHz template. 
Significant correlations are found at the three Galactic latitude cuts at
all frequencies.
The map at 408 MHz is regarded as a good tracer of synchrotron at large scales.
However, at the angular scales left by  the atmospheric filtering in the COSMOSOMAS data processing, the 408 MHz 
template includes  a significant contribution from  extragalactic radiosources 
as can be seen from direct comparison with the NVSS template plotted in  Fig.~\ref{fig-nvss-has},  see also \citep{burigana06}. 
In order to disentangle the contribution of extragalactic radiosources and 
Galactic synchrotron at high Galactic latitude, we performed a simultaneous multi-correlation analysis including the NVSS template. The results are given 
in the lower section of Tab.~\ref{tab-has} and are interpreted as evidence for a rather weak  Galactic synchrotron 
signal  left at our angular scales.. 
Although, the dispersion of the values is high for a precise determination of the synchrotron 
spectral index, the results between 11 GHz and 22.5 GHz are compatible with a 
temperature spectral index of -3, characteristic of  synchrotron emission.

We have also studied the correlation of our maps with a desourced (Ds) Haslam et al. map given by the LAMBDA
archive~\footnote{http://lambda.gsfc.nasa.gov/product/foreground/haslam\_408.cfm}. 
The results improve in the sense that a lower  correlation with NVSS is found, but synchrotron
is still dominated by extragalactic radio sources at our angular scales.  A  very similar conclusion
follows when using 1420 MHz data \citep{reich82}. Finally, we have built a COSMO11 Q map, combining 
the data from the two channels, and performed a  cross-correlation with the Wollaben et al. (2006) Q map at 1420 MHz.
The potential signals are at the level of the statistical errors in the cross-correlation and therefore no conclusion
could be drawn for a polarized Galactic synchrotron signal at 11 GHz. An  upper limit on the polarization
level is set at 2 $\mu$K (68 \% C.L.).

\begin{table*}
\caption{Temperature cross correlation between the COSMOSOMAS and WMAP maps and the
408 MHz map. The first table is for a single correlation analysis. The second one for the
multifit case where NVSS is included in the list of templates. 
The reduction in the amplitude of the  correlation values is due exclusively to the inclusion of
the NVSS template in the fit. Errors are 1 $\sigma$.}
\label{tab-has}
{\tiny
\begin{center}
\begin{tabular}{lcccccccccc}
\hline
Template & ${\rm C11}_1$& ${\rm C11}_2$& C13& C15& C16& WMAP\_K& WMAP\_Ka& WMAP\_Q& WMAP\_V& WMAP\_W \\
\hline
\hline
\multicolumn{11}{c}{$|b|$$>30^\circ$}  \\
\hline
408 MHz & $\mathbf{23.6 \pm 0.9}$ & $\mathbf{24.0 \pm 0.8}$ & $\mathbf{11.9 \pm 0.9}$ & $\mathbf{8.1 \pm 1.2}$ & $4.6 \pm 2.8$ & $\mathbf{3.5 \pm 0.2}$ & $\mathbf{1.5 \pm 0.2}$ & $\mathbf{0.9 \pm 0.2}$ & $\mathbf{0.7 \pm 0.2}$ & $0.1 \pm 0.2$ \\
\hline
\multicolumn{11}{c}{$|b|$$>40^\circ$}  \\
\hline
408 MHz & $\mathbf{25.1 \pm 1.0}$ & $\mathbf{26.3 \pm 0.9}$ & $\mathbf{11.5 \pm 1.1}$ & $\mathbf{7.6 \pm 1.4}$ & $3.5 \pm 3.4$ & $\mathbf{3.9 \pm 0.2}$ & $\mathbf{1.7 \pm 0.2}$ & $\mathbf{1.1 \pm 0.2}$ & $\mathbf{0.9 \pm 0.2}$ & $0.3 \pm 0.2$ \\
\hline
\multicolumn{11}{c}{$|b|$$>50^\circ$}  \\
\hline
408 MHz & $\mathbf{23.1 \pm 1.1}$ & $\mathbf{24.0 \pm 1.0}$ & $\mathbf{9.5 \pm 1.2}$ & $\mathbf{4.3 \pm 1.5}$ & $1.9 \pm 3.7$ & $\mathbf{4.4 \pm 0.3}$ & $\mathbf{2.4 \pm 0.3}$ & $\mathbf{1.6 \pm 0.2}$ & $\mathbf{1.3 \pm 0.2}$ & $\mathbf{1.0 \pm 0.2}$ \\
\hline
\hline
\end{tabular}
\end{center}
}

{\tiny
\begin{center}
\begin{tabular}{lccccccccc}
\hline
Template & ${\rm C11}_1$& ${\rm C11}_2$& C13& C15& C16& WMAP\_K& WMAP\_Ka& WMAP\_Q& WMAP\_W \\
\hline
\hline
\multicolumn{10}{c}{$|b|$$>30^\circ$}  \\
\hline
408 MHz & $\mathbf{2.8 \pm 1.0}$ & $\mathbf{2.8 \pm 0.9}$ & $1.1 \pm 1.1$ & $-1.3 \pm 1.4$ & $-5.4 \pm 3.3$ & $0.0 \pm 0.3$ & $-0.1 \pm 0.3$ & $-0.1 \pm 0.3$ & $-0.4 \pm 0.3$ \\
\hline
\multicolumn{10}{c}{$|b|$$>40^\circ$}  \\
\hline
408 MHz & $\mathbf{5.6 \pm 1.2}$ & $\mathbf{6.0 \pm 1.1}$ & $1.0 \pm 1.3$ & $-2.2 \pm 1.7$ & $-6.2 \pm 4.0$ & $0.1 \pm 0.3$ & $-0.2 \pm 0.3$ & $-0.2 \pm 0.3$ & $-0.5 \pm 0.3$ \\
\hline
\multicolumn{10}{c}{$|b|$$>50^\circ$}  \\
\hline
408 MHz & $\mathbf{4.2 \pm 1.3}$ & $\mathbf{4.7 \pm 1.2}$ & $-0.6 \pm 1.5$ & $\mathbf{-5.0 \pm 1.8}$ & $-5.4 \pm 4.3$ & $0.3 \pm 0.4$ & $0.2 \pm 0.4$ & $-0.1 \pm 0.3$ & $-0.3 \pm 0.4$ \\
\hline
\hline
\end{tabular}
\end{center}
}

\end{table*}

\subsection{Correlations with $H{\alpha}$ data}
\label{sec-halpha}
 Free-free  emission is a likely contributor at our observing frequencies, especially relevant 
when discussing any possible  dust correlated emission.
At high Galactic latitudes, absorption corrections can be ignored at 
first order and we  adopt the combined map of H${\alpha}$ emission provided by 
\citep{finkbeiner03} as an accurate enough tracer of free-free emission.
Tab.~\ref{tbl-H_alpha} gives the temperature
correlation coefficient between COSMOSOMAS and WMAP data 
and this  $H{\alpha}$ map filtered according to the  COSMOSOMAS data processing. 
From the table, we obtain a  conversion from Rayleighs to K of order
40--60 $\mu$K/R for both channels of COSMO11.
Theoretical predictions (Dickinson et al. 2003, Bennet et al. 2003)
for high Galactic latitude yield, assuming a mean electron temperature of
7000-8000 K,  51--55 $\mu $K/R  at 10.9 GHz.
This is compatible with our results.

\begin{table*}
\caption{Values of the temperature correlation in $ \mu $K between
 COSMOSOMAS/WMAP and the  H${\alpha} $ map for different cuts in Galactic
latitude, $b$.  Errors are 1 $\sigma$.}
\label{tbl-H_alpha}
{\tiny
\begin{center}
\begin{tabular}{lccccccccc}
\hline
Template & ${\rm C11}_1$& ${\rm C11}_2$& C13& C15& C16& WMAP\_K& WMAP\_Ka& WMAP\_Q& WMAP\_W \\
\hline
\hline
\multicolumn{10}{c}{$|b|$$>30^\circ$}  \\
\hline
H$\alpha$ & $\mathbf{6.4 \pm 0.9}$ & $\mathbf{6.8 \pm 0.8}$ & $\mathbf{4.1 \pm 0.9}$ & $\mathbf{2.5 \pm 1.2}$ & $-1.7 \pm 2.9$ & $\mathbf{0.9 \pm 0.3}$ & $0.3 \pm 0.3$ & $0.0 \pm 0.2$ & $0.0 \pm 0.2$ \\
\hline
\multicolumn{10}{c}{$|b|$$>40^\circ$}  \\
\hline
H$\alpha$ & $\mathbf{2.7 \pm 1.0}$ & $\mathbf{2.1 \pm 0.9}$ & $1.0 \pm 1.2$ & $\mathbf{4.0 \pm 1.5}$ & $0.2 \pm 3.5$ & $0.4 \pm 0.3$ & $0.3 \pm 0.3$ & $0.1 \pm 0.3$ & $0.0 \pm 0.3$ \\
\hline
\multicolumn{10}{c}{$|b|$$>50^\circ$}  \\
\hline
H$\alpha$ & $\mathbf{4.0 \pm 1.2}$ & $\mathbf{3.2 \pm 1.1}$ & $-1.6 \pm 1.3$ & $2.7 \pm 1.7$ & $-1.8 \pm 3.9$ & $0.2 \pm 0.3$ & $0.1 \pm 0.3$ & $-0.1 \pm 0.3$ & $-0.1 \pm 0.3$ \\
\hline
\hline
\end{tabular}
\end{center}
}

\end{table*}

\subsection{Correlations with dust data}
\label{correl-wmap-dirbe}

The temperature correlations between COSMOSOMAS/WMAP and several dust templates are shown in 
Tab.~\ref{tbl-DUST} for various Galactic latitude cuts. These results are obtained with the 
multifit scheme  described above. We have verified that  they change by  less than 
 $\sim 1 \mu$K   with respect  performing  a simple one to one correlation   
between the COSMOSOMAS and  WMAP data and the  dust templates. Within statistical errors 
the results   are also  independent on the  templates used in the multifit.

\begin{table*}
\caption{Values of the temperature  correlation between
the COSMOSOMAS/WMAP maps and dust maps for different
Galactic latitudes. The values are obtained from a multifit, see text. Errors are 1 $\sigma$.}
\label{tbl-DUST}
{\tiny
\begin{center}
\begin{tabular}{lcccccccccc}
\hline
Template & 1420 MHz& ${\rm C11}_1$& ${\rm C11}_2$& C13& C15& C16& WMAP\_K& WMAP\_Ka& WMAP\_Q& WMAP\_W \\
\hline
\hline
\multicolumn{11}{c}{$|b|$$>30^\circ$}  \\
\hline
$\Lambda 100 $ & $525.1 \pm 569.1$ & $\mathbf{9.1 \pm 0.9}$ & $\mathbf{10.1 \pm 0.8}$ & $\mathbf{4.4 \pm 0.9}$ & $\mathbf{4.9 \pm 1.1}$ & $\mathbf{7.0 \pm 2.7}$ & $\mathbf{2.7 \pm 0.3}$ & $\mathbf{0.7 \pm 0.3}$ & $0.3 \pm 0.2$ & $-0.1 \pm 0.2$ \\
DIRBE08 & $518.0 \pm 578.1$ & $\mathbf{11.4 \pm 0.9}$ & $\mathbf{12.5 \pm 0.8}$ & $\mathbf{5.8 \pm 0.9}$ & $\mathbf{6.3 \pm 1.2}$ & $\mathbf{5.9 \pm 2.9}$ & $\mathbf{2.8 \pm 0.3}$ & $\mathbf{0.7 \pm 0.3}$ & $0.3 \pm 0.2$ & $-0.2 \pm 0.2$ \\
DIRBE10 & $616.0 \pm 566.1$ & $\mathbf{9.7 \pm 0.9}$ & $\mathbf{11.3 \pm 0.8}$ & $\mathbf{3.7 \pm 0.9}$ & $1.9 \pm 1.2$ & $5.2 \pm 2.9$ & $\mathbf{2.1 \pm 0.3}$ & $0.5 \pm 0.3$ & $0.2 \pm 0.2$ & $-0.3 \pm 0.2$ \\
\hline
\multicolumn{11}{c}{$|b|$$>40^\circ$}  \\
\hline
$\Lambda 100 $ & $-617.0 \pm 663.0$ & $\mathbf{6.2 \pm 1.0}$ & $\mathbf{7.2 \pm 1.0}$ & $0.4 \pm 1.2$ & $\mathbf{3.4 \pm 1.4}$ & $1.5 \pm 3.5$ & $\mathbf{1.5 \pm 0.3}$ & $\mathbf{0.6 \pm 0.3}$ & $0.3 \pm 0.3$ & $0.1 \pm 0.3$ \\
DIRBE08 & $-955.0 \pm 663.0$ & $\mathbf{6.1 \pm 1.1}$ & $\mathbf{7.4 \pm 1.0}$ & $1.0 \pm 1.2$ & $2.3 \pm 1.4$ & $0.0 \pm 3.5$ & $\mathbf{1.2 \pm 0.3}$ & $0.5 \pm 0.3$ & $0.2 \pm 0.3$ & $0.0 \pm 0.3$ \\
DIRBE10 & $-314.0 \pm 657.1$ & $\mathbf{4.7 \pm 1.0}$ & $\mathbf{6.2 \pm 0.9}$ & $1.4 \pm 1.2$ & $-0.7 \pm 1.5$ & $-0.6 \pm 3.5$ & $\mathbf{0.8 \pm 0.3}$ & $0.3 \pm 0.3$ & $0.2 \pm 0.3$ & $-0.1 \pm 0.3$ \\
\hline
\multicolumn{11}{c}{$|b|$$>50^\circ$}  \\
\hline
$\Lambda 100 $ & $\mathbf{-1487.0 \pm 732.1}$ & $\mathbf{2.6 \pm 1.2}$ & $1.8 \pm 1.1$ & $\mathbf{2.6 \pm 1.3}$ & $\mathbf{3.6 \pm 1.6}$ & $-2.8 \pm 4.0$ & $\mathbf{1.4 \pm 0.3}$ & $0.5 \pm 0.3$ & $0.3 \pm 0.3$ & $0.2 \pm 0.3$ \\
DIRBE08 & $\mathbf{-1660.0 \pm 731.1}$ & $1.6 \pm 1.2$ & $1.6 \pm 1.1$ & $\mathbf{2.6 \pm 1.3}$ & $2.0 \pm 1.6$ & $-5.0 \pm 4.0$ & $\mathbf{0.9 \pm 0.3}$ & $0.4 \pm 0.3$ & $0.2 \pm 0.3$ & $0.0 \pm 0.3$ \\
DIRBE10 & $-651.1 \pm 723.0$ & $\mathbf{2.8 \pm 1.1}$ & $\mathbf{4.4 \pm 1.0}$ & $\mathbf{3.6 \pm 1.3}$ & $0.2 \pm 1.6$ & $-5.0 \pm 3.9$ & $\mathbf{0.8 \pm 0.3}$ & $0.2 \pm 0.3$ & $0.2 \pm 0.3$ & $-0.1 \pm 0.3$ \\
\hline
\hline
\end{tabular}
\end{center}
}

\end{table*}

For $|b|>30^{\circ}$, the most significant correlations  are
obtained for both COSMO11 channels, with a signal to noise ratio
greater than 10, and for the WMAP K channel, with a signal to noise
around 7--9. Marginal detections are also found with the K$a$
channel. No significant detection is found at the higher frequency
channels of WMAP. The channels of COSMO15 also exhibit some clear
detections, although for some templates these are less significant.
At higher Galactic latitudes the amplitude of the correlated signal
decreases and for $|b| >50^{\circ} $ the detections are below the
3$\sigma$ level. Overall, we find clear evidence for a dust correlated
high Galactic latitude microwave emission in the frequency range 11-33 GHz.

We note that the two COSMO11 channels present a small systematic
difference in the values of the correlations at $|b|>30^{\circ}$ and
$|b|>40^{\circ}$, of order 1-1.5 $\mu$K, similar to the
statistical error. In principle, this could be due to a significant
level of polarization in the dust-correlated signal but more sensitive
maps are required in order to establish whether this is a reliable difference.

The dust-correlated signal detected in Tab.~\ref{tbl-DUST} presents a
spectral index between 11 and 33 GHz of -2.1$\pm$0.2, incompatible with 
synchrotron emission ($\sim -3$) but consistent with free-free emission.
However, the remarkable   flattening  found  in the  frequency range of the
 COSMOSOMAS experiment cannot be understood in terms of free-free emission 
properly, suggesting an improper tracing  of the free-free emission at high
 Galactic latitude by the H$\alpha$ template used
in the multifit analysis. This would be the case in regions of high dust 
emission where dust also causes  high extinction in the visual.
 A careful study of the regions that mostly contribute to the
dust-correlated signal in the COSMO11 data at $|b|>30^{\circ}$ reveals
areas  with a very strong emission in the DIRBE maps which indeed
are not H$\alpha$ emitters. Dust extinction in these regions would
cause that free-free emission would not be traced by the H$\alpha$
template affecting to the correlation values listed in Tab.~\ref{tbl-DUST}. 
In order to explore if this is the case, we identified
regions with very strong dust emission at high Galactic latitudes
and performed a new correlation study by extending the mask to include
those pixels whith a significant correlated emission with the COSMOSOMAS
maps. For definiteness, we extended the mask to  those pixels where the amplitude of
the normalised covariance between the channel 2 of COSMOSOMAS (the one
with the best sensitivity) and DIRBE08 (i.e. $<t_{\rm C2} t_{\rm D08}>
/ (\sigma(t_{\rm C2}) \sigma(t_{\rm D08}))$) was greater than $5$. Adjacent pixels
were also masked.
In total, only a few regions of size $1-5$ square degrees were
removed from the analysis and a new multifit correlation analysis was performed
 The results are listed in Tab.~\ref{tab-dust-sigma}.  We
 find a decrease in the amplitude of the dust- correlated
signal at any frequency with respect to the case in
Tab.~\ref{tbl-DUST} but the correlations at  $|b|>30^{\circ}$
 are  very significant for COSMO11 and WMAP\_K and significant for the other
COSMOSOMAS channels. There is also a  trend to lower amplitudes as we move to 
high Galactic latitude. In order to see if this dust-correlated microwave emission
could be associated with synchrotron, free-free or a different emission process
we conducted  a detailed analysis of the DIRBE08 correlation values listed in
Tab.~\ref{tab-dust-sigma} for $|b|>30^{\circ}$.

\begin{table*}
\caption{Temperature correlations between
the COSMOSOMAS and WMAP maps and dust maps for different Galactic
latitudes. The values are obtained extending the mask to the
particular regions where there is a high correlation between
COSMOSOMAS and DIRBE08, see text. Errors are 1 $\sigma$.}
\label{tab-dust-sigma}
{\tiny
\begin{center}
\begin{tabular}{lcccccccccc}
\hline
Template & 1420 MHz& C1& C2& C13& C15& C17& WMAP\_K& WMAP\_Ka& WMAP\_Q& WMAP\_W \\
\hline
\hline
\multicolumn{11}{c}{$|b|$$>30^\circ$}  \\
\hline
$\Lambda 100 $ & $96.1 \pm 571.0$ & $\mathbf{4.7 \pm 0.9}$ & $\mathbf{5.9 \pm 0.8}$ & $\mathbf{2.4 \pm 0.9}$ & $\mathbf{2.3 \pm 1.1}$ & $4.5 \pm 2.7$ & $\mathbf{1.9 \pm 0.3}$ & $0.4 \pm 0.3$ & $0.2 \pm 0.2$ & $-0.1 \pm 0.2$ \\
DIRBE08 & $-139.1 \pm 573.1$ & $\mathbf{4.6 \pm 0.9}$ & $\mathbf{5.8 \pm 0.8}$ & $\mathbf{2.8 \pm 0.9}$ & $\mathbf{3.5 \pm 1.2}$ & $2.5 \pm 2.9$ & $\mathbf{1.8 \pm 0.3}$ & $0.4 \pm 0.3$ & $0.2 \pm 0.2$ & $-0.2 \pm 0.2$ \\
DIRBE10 & $115.0 \pm 568.1$ & $\mathbf{4.8 \pm 0.9}$ & $\mathbf{6.6 \pm 0.8}$ & $1.2 \pm 0.9$ & $-0.6 \pm 1.2$ & $2.8 \pm 2.9$ & $\mathbf{1.3 \pm 0.3}$ & $0.3 \pm 0.3$ & $0.1 \pm 0.2$ & $-0.3 \pm 0.2$ \\
\hline
\multicolumn{11}{c}{$|b|$$>40^\circ$}  \\
\hline
$\Lambda 100 $ & $-962.0 \pm 663.1$ & $\mathbf{3.5 \pm 1.0}$ & $\mathbf{3.7 \pm 0.9}$ & $-0.4 \pm 1.2$ & $1.9 \pm 1.5$ & $0.6 \pm 3.5$ & $\mathbf{1.4 \pm 0.3}$ & $\mathbf{0.6 \pm 0.3}$ & $0.3 \pm 0.3$ & $0.2 \pm 0.3$ \\
DIRBE08 & $-1252.0 \pm 663.0$ & $\mathbf{3.4 \pm 1.1}$ & $\mathbf{3.9 \pm 1.0}$ & $0.3 \pm 1.2$ & $0.8 \pm 1.5$ & $-1.2 \pm 3.5$ & $\mathbf{1.0 \pm 0.3}$ & $0.4 \pm 0.3$ & $0.2 \pm 0.3$ & $0.1 \pm 0.3$ \\
DIRBE10 & $-443.1 \pm 659.0$ & $\mathbf{3.5 \pm 1.0}$ & $\mathbf{4.7 \pm 0.9}$ & $1.1 \pm 1.2$ & $-1.4 \pm 1.5$ & $-1.0 \pm 3.6$ & $\mathbf{0.7 \pm 0.3}$ & $0.2 \pm 0.3$ & $0.1 \pm 0.3$ & $-0.1 \pm 0.3$ \\
\hline
\multicolumn{11}{c}{$|b|$$>50^\circ$}  \\
\hline
$\Lambda 100 $ & $\mathbf{-1541.0 \pm 733.0}$ & $1.8 \pm 1.2$ & $0.7 \pm 1.1$ & $2.5 \pm 1.3$ & $\mathbf{3.5 \pm 1.6}$ & $-2.8 \pm 4.0$ & $\mathbf{1.3 \pm 0.3}$ & $0.5 \pm 0.3$ & $0.3 \pm 0.3$ & $0.2 \pm 0.3$ \\
DIRBE08 & $\mathbf{-1710.0 \pm 732.0}$ & $0.7 \pm 1.2$ & $0.4 \pm 1.1$ & $2.5 \pm 1.3$ & $2.0 \pm 1.6$ & $-4.9 \pm 4.0$ & $\mathbf{0.9 \pm 0.3}$ & $0.4 \pm 0.3$ & $0.2 \pm 0.3$ & $0.0 \pm 0.3$ \\
DIRBE10 & $-656.1 \pm 723.1$ & $\mathbf{2.7 \pm 1.1}$ & $\mathbf{4.3 \pm 1.0}$ & $\mathbf{3.6 \pm 1.3}$ & $0.2 \pm 1.6$ & $-4.8 \pm 3.9$ & $\mathbf{0.8 \pm 0.3}$ & $0.1 \pm 0.3$ & $0.2 \pm 0.3$ & $-0.1 \pm 0.3$ \\
\hline
\hline
\end{tabular}
\end{center}
}

\end{table*}

%
A maximum likelihood approach has been used to find the best-fit
parameters within three different models.  First, we considered a
single power law with two free parameters ($f(\nu) = A (\nu/22~{\rm
GHz})^{p}$) as a fitting function; second, we considered a
two-component fit, in which one component is given by a power-law, and
the other is described by the standard combination of Draine \&
Lazarian models (CNM, WNM and WIM, with relative amplitudes of 0.43,
0.43 and 0.14, respectively). Finally, we considered a
phenomenological approach and tried to fit the data with a single
Gaussian law given by $ f(\nu) = A \, \exp{(- (\nu-\nu_0)^2}/(2
\sigma^2)) $ in order to mimic the behaviour of a general
spinning-dust model.

The best fit is obtained in the third case, with a reduced value
of $\chi^2 /{\rm dof} = 0.72$ (probability of 63\%), although the
other two cases show similar values for the goodness-of-fit, obtaining
$1.05$ (39.1\%) and $0.93$ (46.9\%) for model 1 and 2,  respectively.
We note that in the case of model 1 the index of the power-law ($p=0.1\pm0.2$))
rules out synchrotron emission but is compatible  with free-free.

\begin{figure}
\begin{center}
\includegraphics[height=7cm,width=9cm,angle=0]{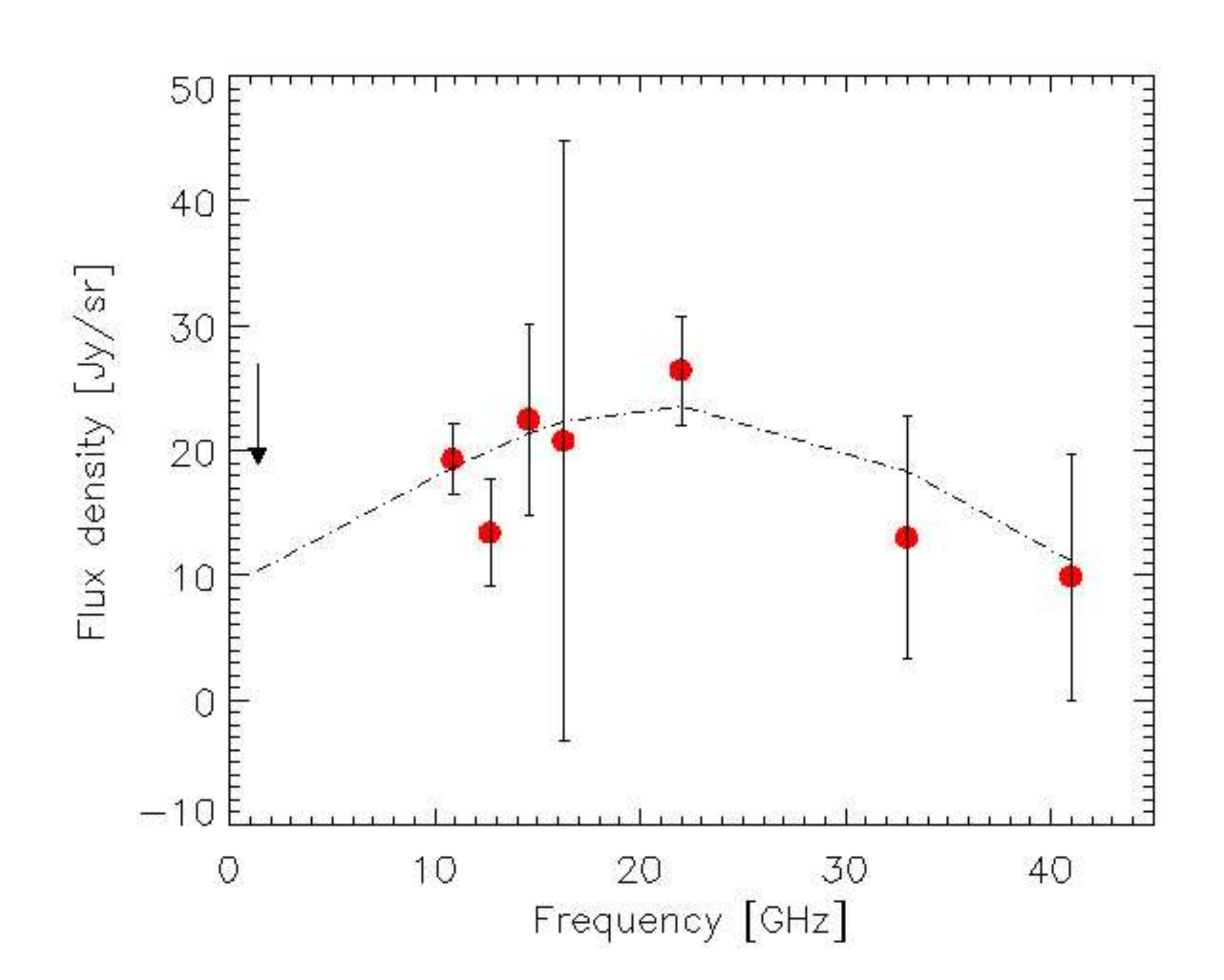}
\caption{Correlation between COSMOSOMAS, WMAP channels and the DIRBE08 map in
flux density units (Jy/sr) for $|b|>30^{\circ}$
after masking some localized regions, see text for details.
Overplotted is the best fit model of a Gaussian model with parameters
$A= 23.6 $Jy/sr, $\nu_0 = 21.7 $ GHz and $ \sigma = 15.8$ GHz.}
\label{fig-flux_density}
\end{center}
\end{figure}

For the third model, the best fit parameters (obtained from the
marginalized likelihood function) are $A =
23.5^{+3.9}_{-3.2}$~Jy/sr, $\nu_0 = 21.7^{+3.8}_{-3.7}$~GHz and
$\sigma =15.8^{+4.0}_{-3.4}$~GHz. In Fig.~\ref{fig-flux_density} we plot 
the data of Tab.~\ref{tab-dust-sigma} and this model.

This ``blind'' best-fit function peaks at a slightly lower frequency
than the considered Draine \& Lazarian model, and has a slightly
broader width. In the case of interpreting this anomalous signal in
terms of a spinning dust model, this would have an impact on the
actual size distribution of the grains responsible for the emission.

In summary, these results appear to support the existence of a 
spinning dust microwave
emission process at high Galactic latitudes. In order to disentangle
whether this anomalous microwave emission is associated to well
defined regions as it is the case of the Perseus molecular complex
(Watson et al. 2005), or to a diffuse large-scale Galactic component,
 maps of higher sensitivity and resolution are required in the
frequency range 10-30 GHz.

\section{Conclusions}
 
We have presented new results of the COSMOSOMAS experiment at 11 GHz
for a sky coverage of ca. 6500 sq. deg. 
A cross-correlation analysis using this new dataset,
previous COSMOSOMAS data in the frequency range 12-17 GHz, the 3rd year
WMAP data and  several foreground templates yields the following results
at high Galactic latitude
($|b|>30^{\circ}$):
 \begin{itemize}
 \item 
The presence of a CMB component of amplitude  27$\pm$2 $\mu$K in the COSMOSOMAS
channels  in agreement with the expected CMB fluctuation after removal of 
the Fourier terms implied in the atmospheric filtering. 

\item 
For the angular scales of our experiment,
unresolved extragalactic sources are found to be the dominant
foreground at 11 GHz.  The measured contributions of unresolved
radiosources in the range 11--23 GHz  are in 
very good agreement with predictions by de Zotti et al. (2005).

\item
The  synchrotron component at high Galactic latitude  is found with an amplitude of 3-6 $\mu$K at 11 GHz, 
and decreases at higher frequency with a temperature spectral index  compatible with  -3. 

\item
The cross-correlation with the  H$_{\alpha}$ map gives an amplitude  of  3-6 $\mu$K at 11 GHz. 
 Correlations with higher frequency COSMOSOMAS and WMAP data
verify a temperature spectral index consistent with -2, characteristic of 
free-free emission.

\item A dust correlated emission is detected in each of the COSMOSOMAS
channels at  $|b| >30^{\circ} $. The amplitude of the signal ranges from 10-12 $\mu$K at 11
GHz down to 4-7 $\mu$K in the 12-17 GHz and 2.1-2.8 $\mu$K at 22
GHz. The Galactic latitude dependence supports a Galactic origin for
this signal.  An important fraction of this correlated signal at 11
GHz comes from regions of high dust emission where free-free emission
is not well traced by the H$\alpha$ template due to extinction. After
masking those regions the remaining dust correlated signal detectable
in the frequency range 11-33 GHz - of order 5-6 $\mu$K at 11 GHz - 
shows a clear flattening, which is not compatible with the classical
spectral index of synchrotron emission. This correlated signal can be
 described by models resembling spinning dust emission,
with an associated flux density peaking around 20~GHz.
 \end{itemize}

\section*{ \bf Acknowledgements}
We are thankful to L. Toffolatti and  J. Gonz\'alez-Bueno for very useful
discussions on unresolved radiosource contributions at the frequencies
of COSMOSOMAS. We also thank J.L. Salazar and the personnel of the IAC engineering
division, instrument maintenance  and operations of Teide Observatory  for their 
support to the COSMOSOMAS experiment.  This research has been partially
funded by project AYA 2001-1657 and AYA 2005-06453 of the Spanish Ministry 
of Education and Science.
\onecolumn
\twocolumn

\bibliographystyle{mn2e}

\label{lastpage}

\end{document}